\documentclass[aps,notitlepage,twocolumn,nofootinbib,superscriptaddress,longbibliography]{revtex4-2}
\usepackage[dvipsnames]{xcolor}
\usepackage{framed}
\definecolor{shadecolor}{rgb}{0.9,0.9,0.9}

\usepackage{mathtools}
\usepackage{amsmath}
\usepackage[shortlabels]{enumitem}
\usepackage{graphicx,epic,eepic,epsfig,amsmath,latexsym,amssymb,verbatim,color}
 
\usepackage{amsfonts}       % blackboard math symbols
\usepackage{nicefrac}       % compact symbols for 1/2, etc.

\usepackage{amsmath}
\usepackage{bbm}

\usepackage{float}
\usepackage{tikz}
\usetikzlibrary{chains}
\usetikzlibrary{fit}
\usepackage{pgflibraryarrows}		%optional
\usepackage{pgflibrarysnakes}		%optional

\usepackage{epsfig}
\usetikzlibrary{shapes.symbols,patterns} % for source symbols
\usepackage{pgfplots}

\usepackage[strict]{changepage}
\usepackage{hyperref}
\hypersetup{colorlinks=true,citecolor=blue,linkcolor=blue,filecolor=blue,urlcolor=blue,breaklinks=true}

\usepackage[marginal]{footmisc}
\usepackage{url}
\usepackage{theorem}

\newtheorem{proposition}{Proposition}
\newtheorem{lemma}[proposition]{Lemma}

\def\squareforqed{\hbox{\rlap{$\sqcap$}$\sqcup$}}
\def\qed{\ifmmode\squareforqed\else{\unskip\nobreak\hfil
\penalty50\hskip1em\null\nobreak\hfil\squareforqed
\parfillskip=0pt\finalhyphendemerits=0\endgraf}\fi}
\def\endenv{\ifmmode\;\else{\unskip\nobreak\hfil
\penalty50\hskip1em\null\nobreak\hfil\;
\parfillskip=0pt\finalhyphendemerits=0\endgraf}\fi}

\newcounter{remark}

\newcounter{example}

\mathchardef\ordinarycolon\mathcode`\:
\mathcode`\:=\string"8000
\def\vcentcolon{\mathrel{\mathop\ordinarycolon}}
\begingroup \catcode`\:=\active
  \lowercase{\endgroup
  \let :\vcentcolon
  }

\usepackage{cleveref}
\usepackage{graphicx}
\usepackage{xcolor}

\RequirePackage[framemethod=default]{mdframed}
\newmdenv[skipabove=7pt,
skipbelow=7pt,
backgroundcolor=darkblue!15,
innerleftmargin=5pt,
innerrightmargin=5pt,
innertopmargin=5pt,
leftmargin=0cm,
rightmargin=0cm,
innerbottommargin=5pt,
linewidth=1pt]{tBox}

\newmdenv[skipabove=7pt,
skipbelow=7pt,
backgroundcolor=red!15,
innerleftmargin=5pt,
innerrightmargin=5pt,
innertopmargin=5pt,
leftmargin=0cm,
rightmargin=0cm,
innerbottommargin=5pt,
linewidth=1pt]{rBox}

\newmdenv[skipabove=7pt,
skipbelow=7pt,
backgroundcolor=blue2!25,
innerleftmargin=5pt,
innerrightmargin=5pt,
innertopmargin=5pt,
leftmargin=0cm,
rightmargin=0cm,
innerbottommargin=5pt,
linewidth=1pt]{dBox}
\newmdenv[skipabove=7pt,
skipbelow=7pt,
backgroundcolor=darkkblue!15,
innerleftmargin=5pt,
innerrightmargin=5pt,
innertopmargin=5pt,
leftmargin=0cm,
rightmargin=0cm,
innerbottommargin=5pt,
linewidth=1pt]{sBox}
\definecolor{darkblue}{RGB}{0,76,156}
\definecolor{darkkblue}{RGB}{0,0,153}
\definecolor{blue2}{RGB}{102,178,255}
\definecolor{darkred}{RGB}{195,0,0}
\newcommand{\nc}{\newcommand}
\nc{\rnc}{\renewcommand}
\nc{\lbar}[1]{\overline{#1}}
\nc{\bra}[1]{\langle#1|}
\nc{\ket}[1]{|#1\rangle}
\nc{\ketbra}[2]{|#1\rangle\!\langle#2|}
\nc{\braket}[2]{\langle#1|#2\rangle}

\nc{\proj}[1]{| #1\rangle\!\langle #1 |}
\nc{\avg}[1]{\langle#1\rangle}
\nc{\smfrac}[2]{\mbox{$\frac{#1}{#2}$}}
\nc{\tr}{\operatorname{Tr}}
\nc{\ox}{\otimes}
\nc{\dg}{\dagger}
\nc{\dn}{\downarrow}
\nc{\cA}{{\cal A}}
\nc{\cB}{{\cal B}}
\nc{\cC}{{\cal C}}
\nc{\cD}{{\cal D}}
\nc{\cE}{{\cal E}}
\nc{\cF}{{\cal F}}
\nc{\cG}{{\cal G}}
\nc{\cH}{{\cal H}}
\nc{\cI}{{\cal I}}
\nc{\cJ}{{\cal J}}
\nc{\cK}{{\cal K}}
\nc{\cL}{{\cal L}}
\nc{\cM}{{\cal M}}
\nc{\cN}{{\cal N}}
\nc{\cO}{{\cal O}}
\nc{\cP}{{\cal P}}
\nc{\cQ}{{\cal Q}}
\nc{\cR}{{\cal R}}
\nc{\cS}{{\cal S}}
\nc{\cT}{{\cal T}}
\nc{\cU}{{\cal U}}
\nc{\cV}{{\cal V}}
\nc{\cX}{{\cal X}}
\nc{\cY}{{\cal Y}}
\nc{\cZ}{{\cal Z}}
\nc{\cW}{{\cal W}}
\nc{\csupp}{{\operatorname{csupp}}}
\nc{\qsupp}{{\operatorname{qsupp}}}
\nc{\var}{{\operatorname{var}}}
\nc{\rar}{\rightarrow}
\nc{\lrar}{\longrightarrow}
\nc{\polylog}{{\operatorname{polylog}}}
\nc{\wt}{{\operatorname{wt}}}
\nc{\av}[1]{{\left\langle {#1} \right\rangle}}
\nc{\supp}{{\operatorname{supp}}}

\nc{\argmin}{{\operatorname{argmin}}}

\def\x{\xi}

\nc{\RR}{{{\mathbb R}}}
\nc{\CC}{{{\mathbb C}}}
\nc{\FF}{{{\mathbb F}}}
\nc{\NN}{{{\mathbb N}}}
\nc{\ZZ}{{{\mathbb Z}}}
\nc{\PP}{{{\mathbb P}}}
\nc{\QQ}{{{\mathbb Q}}}
\nc{\UU}{{{\mathbb U}}}
\nc{\EE}{{{\mathbb E}}}
\nc{\id}{{\operatorname{id}}}

\nc{\CHSH}{{\operatorname{CHSH}}}

\nc{\be}{\begin{equation}}
\nc{\ee}{{\end{equation}}}
\nc{\bea}{\begin{eqnarray}}
\nc{\eea}{\end{eqnarray}}
\nc{\<}{\langle}
\rnc{\>}{\rangle}
\nc{\rU}{\mbox{U}}

\nc{\ob}[1]{#1}

\nc{\SEP}{{\text{\rm SEP}}}
\nc{\NS}{{\text{\rm NS}}}
\nc{\LOCC}{{\text{\rm LOCC}}}
\nc{\PPT}{{\text{\rm PPT}}}
\nc{\EXT}{{\text{\rm EXT}}}
\nc{\Sym}{{\operatorname{Sym}}}

\nc{\ERLO}{{E_{\text{r,LO}}}}
\nc{\ERLOCC}{{E_{\text{r,LOCC}}}}
\nc{\ERPPT}{{E_{\text{r,PPT}}}}
\nc{\ERLOCCinfty}{{E^{\infty}_{\text{r,LOCC}}}}
\nc{\Aram}{{\operatorname{\sf A}}}

\usepackage{tikz}
\usepackage{hyperref}
\hypersetup{colorlinks=true,citecolor=blue,linkcolor=blue,filecolor=blue,urlcolor=blue,breaklinks=true}

\makeatletter
\def\grd@save@target#1{%
  \def\grd@target{#1}}
\def\grd@save@start#1{%
  \def\grd@start{#1}}
\tikzset{
  grid with coordinates/.style={
    to path={%
      \pgfextra{%
        \edef\grd@@target{(\tikztotarget)}%
        \tikz@scan@one@point\grd@save@target\grd@@target\relax
        \edef\grd@@start{(\tikztostart)}%
        \tikz@scan@one@point\grd@save@start\grd@@start\relax
        \draw[minor help lines,magenta] (\tikztostart) grid (\tikztotarget);
        \draw[major help lines] (\tikztostart) grid (\tikztotarget);
        \grd@start
        \pgfmathsetmacro{\grd@xa}{\the\pgf@x/1cm}
        \pgfmathsetmacro{\grd@ya}{\the\pgf@y/1cm}
        \grd@target
        \pgfmathsetmacro{\grd@xb}{\the\pgf@x/1cm}
        \pgfmathsetmacro{\grd@yb}{\the\pgf@y/1cm}
        \pgfmathsetmacro{\grd@xc}{\grd@xa + \pgfkeysvalueof{/tikz/grid with coordinates/major step}}
        \pgfmathsetmacro{\grd@yc}{\grd@ya + \pgfkeysvalueof{/tikz/grid with coordinates/major step}}
        \foreach \x in {\grd@xa,\grd@xc,...,\grd@xb}
        \node[anchor=north] at (\x,\grd@ya) {\pgfmathprintnumber{\x}};
        \foreach \y in {\grd@ya,\grd@yc,...,\grd@yb}
        \node[anchor=east] at (\grd@xa,\y) {\pgfmathprintnumber{\y}};
      }
    }
  },
  minor help lines/.style={
    help lines,
    step=\pgfkeysvalueof{/tikz/grid with coordinates/minor step}
  },
  major help lines/.style={
    help lines,
    line width=\pgfkeysvalueof{/tikz/grid with coordinates/major line width},
    step=\pgfkeysvalueof{/tikz/grid with coordinates/major step}
  },
  grid with coordinates/.cd,
  minor step/.initial=.2,
  major step/.initial=1,
  major line width/.initial=2pt,
}
\makeatother

\usepackage{thmtools}
\usepackage{thm-restate}
\usepackage{etoolbox}
\makeatletter
\def\problem@s{}
\newcounter{problems@cnt}

\newcommand{\allproblems}{\problem@s}
\makeatother

\usepackage[normalem]{ulem}
\usepackage{algorithmic}
\usepackage[ruled, vlined, linesnumbered]{algorithm2e}
\usepackage{times}
\usepackage{amsmath,amsfonts,amssymb,graphicx,mathtools,bm,tcolorbox,relsize}
\usepackage[utf8]{inputenc}
\usepackage[T1]{fontenc}
\usepackage[qm]{qcircuit}
\usepackage{braket}
\usepackage{natbib}
\usepackage{multirow}
\usepackage{booktabs}
\usepackage{tabularx}
\usepackage{array}
\usepackage{optidef}

\pgfplotsset{compat=1.18}
\definecolor{defcolor}{RGB}{140, 149, 170}
\definecolor{thmcolor}{RGB}{127, 139, 117}
\definecolor{conjcolor}{RGB}{154, 85, 87}
\tcbuselibrary{theorems}
\tcbset{
    defstyle/.style={colback=defcolor!5,
    colframe=defcolor!100!white,
    fonttitle=\bfseries,},
    thmstyle/.style={colback=thmcolor!5,
    colframe=thmcolor!100!white,
    fonttitle=\bfseries,},
    conjstyle/.style={colback=conjcolor!5,
    colframe=conjcolor!100!white,
    fonttitle=\bfseries,},
}
\usepackage[scr=boondox]{mathalpha}

\allowdisplaybreaks

\begin{document}
\title{Virtual quantum neural networks}

\author{Benchi Zhao}
\affiliation{QICI Quantum Information and Computation Initiative, School of Computing and Data Science, The University of Hong Kong, Pokfulam Road, Hong Kong}

\author{Xuanqiang Zhao}
\affiliation{QICI Quantum Information and Computation Initiative, School of Computing and Data Science, The University of Hong Kong, Pokfulam Road, Hong Kong}

\author{Yinan Li}
% \email{yinan.li@whu.edu.cn}
\affiliation{School of Artificial Intelligence, Wuhan University, Wuhan, China}
\affiliation{Hubei Center for Applied Mathematics, Wuhan, China}
\affiliation{Hubei Key Laboratory of Computational Science, Wuhan, China}
\affiliation{Wuhan Institute of Quantum Technology, Wuhan, China}

\author{Yingzhou Li}
\email{yingzhouli@fudan.edu.cn}
\affiliation{School of Mathematical Sciences, Shanghai Key Laboratory for Contemporary Applied Mathematics, Fudan University}
\affiliation{Key Laboratory of Computational Physical Sciences, Ministry of Education}

\author{Giulio Chiribella}
\email{giulio@hku.hk}
\affiliation{QICI Quantum Information and Computation Initiative, School of Computing and Data Science, The University of Hong Kong, Pokfulam Road, Hong Kong}
\affiliation{Quantum Group, Department of Computer Science, University of Oxford, Wolfson Building, Parks Road, Oxford, OX1 3QD, United Kingdom}
\affiliation{Perimeter Institute for Theoretical Physics, 31 Caroline Street North, Waterloo, Ontario, Canada}
%%%%%%%%%%%%%%%%%%%%%%%%%%%%%%%%%%%%%%%%%%%%%%%%%%%%%%%%%%%%
%%%%%%%%%%%%%%%%%%%%%%%%%%%%%%%%%%%%%%%%%%%%%%%%%%%%%%%%%%%%
\begin{abstract}
Quantum neural networks are a prominent  model of quantum machine learning.   Their training  consists in the minimization of  a given loss function over a   parametrized  family of quantum circuits, mathematically described by unitary operators, or, more generally, completely positive linear  maps.  %The restriction to this set of operators/maps, however, is not essential:  when classical sampling and data processing are exploited,  a larger optimization space becomes accessible,  corresponding to a larger set of Hermitian-preserving maps.
In this work, we extend the notion of quantum neural network,  using random sampling and classical data processing to enlarge the optimization space in a way that includes linear combinations of completely positive maps.
%in a way that includes  all possible linear combinations of the maps corresponding to the sampled circuits.
%including all possible linear combinations of the   to  the broader set of   Hermitian-preserving maps.
%also known as virtual quantum operations.
Our extended model, called  {\it virtual quantum neural networks}, leverages its enlarged  optimization space to achieve increased expressivity  %without exacerbating barren plateaus
and  improved  noise robustness. These benefits are illustrated in three representative  tasks:  quantum error mitigation, binary classification, and  estimation of ground-state energies. Overall, virtual quantum neural networks offer a flexible learning paradigm that  expands the space of achievable computations and strengthens the applications   of near-term quantum hardware.

\end{abstract}

% \date{\today}
\maketitle

%%%%%%%%%%%%%%%%%%%%%%%%%%%%%%%%%%%%%%%%%%%%%%%%%%%%%%%%%%%%
%%%%%%%%%%%%%%%%%%%%%%%%%%%%%%%%%%%%%%%%%%%%%%%%%%%%%%%%%%%%
\section{Introduction}

The present stage of quantum computing is known as the noisy intermediate-scale quantum (NISQ) era ~\cite{preskill2018quantum}. It this stage,   noisy quantum computations of moderate depth can be implemented on a few hundred qubits, starting to probe a new regime beyond the reach of classical numerical simulations.  Exploiting these capabilities, several applications of NISQ quantum processors have been identified, including  optimization~\cite{farhi2014quantum,harrigan2021quantum},  chemistry and material science~\cite{mcardle2020quantum, google2020hartree, nakanishi2019subspace}, and cryptography~\cite{bennett2014quantum,ekert1991quantum}, a broad class of NISQ algorithms has been designed~\cite{fujii2022deep,yuan2019theory, zhao2025variational, endo2020variational,zhao2021practical, chen2023near, chen2021variational}.

 %at the intersection of quantum computing and  machine learning. which aims to leverage the expressive power of high-dimensional Hilbert spaces and intrinsic quantum correlations to enhance learning tasks.

  A notable approach to quantum computing in the NISQ era is provided by quantum neural networks (QNNs)~\cite{wan2017quantum,farhi2018classification,mitarai2018quantum,schuld2020circuit,cong2019quantum} (see also \cite{du2025quantum} for a review in the broader context of quantum machine learning).
 In a QNN, an achievable family of quantum circuits is specified by a set of  parameters, which are optimized according to a  loss function evaluated on classical data obtained from   measurements. This quantum–classical optimization loop  is appealing for near-term applications because it offers adaptability to hardware limitations while retaining sufficient expressivity for realizing non-trivial quantum computations.   It provides a prominent model of  quantum machine learning (QML)~\cite{biamonte2017quantum, PhysRevLett.113.130503} and  underpins  all variational quantum algorithms (VQAs)~\cite{cerezo2021variational},  supporting QML primitives such as  quantum classifiers~\cite{li2022recent,grant2018hierarchical}, data compression~\cite{romero2017quantum, cao2021noise, bondarenko2020quantum}, and generative models~\cite{gao2018quantum, dallaire2018quantum}. %From this perspective, QML models can be viewed as instances of VQAs tailored to learning tasks, where the PQC acts as a trainable model and the classical optimizer updates its parameters based on measurement outcomes.

%%%%%%%%%%%%%%%%%%%%%%%%%%%%%%%%%%%%%%%%%%%%%%%%%%%%%%%%%%%%%%%%%%%%%%%%%%%%%%%%%%%%%%%%%%%%%%

%over multiple families of parametrized quantum circuits.

Technically, a   QNN is described by a parametrized family of unitary operators, or, more generally,  a parametrized family of completely positive trace-preserving (CPTP) maps.  The minimization of the loss function is then performed on the family of CPTP maps reachable by varying the QNN parameters. This framework, however, entails a fundamental limitation,  since the optimization is by default constrained to a subset of CPTP maps  achievable on the given quantum hardware.  This limitation restricts the expressivity of the model, typically restricting it to computations represented by shallow circuits,  and in turn affects its practical applications. 

In this paper, we develop a more general model that lifts the above limitations by   sampling over multiple QNNs and by classically postprocessing the measurement outcomes obtained in different runs. The combined use of random sampling and classical postprocessing allows us to access a broader set of maps, consisting of all possible linear combinations of the CPTP maps achievable by the sampled  QNNs.   This extension  provides a larger optimization  space, increasing the expressivity of the model and potentially  unlocking new quantum advantages.

\begin{figure*}[t]
    \centering
    \includegraphics[width=\linewidth]{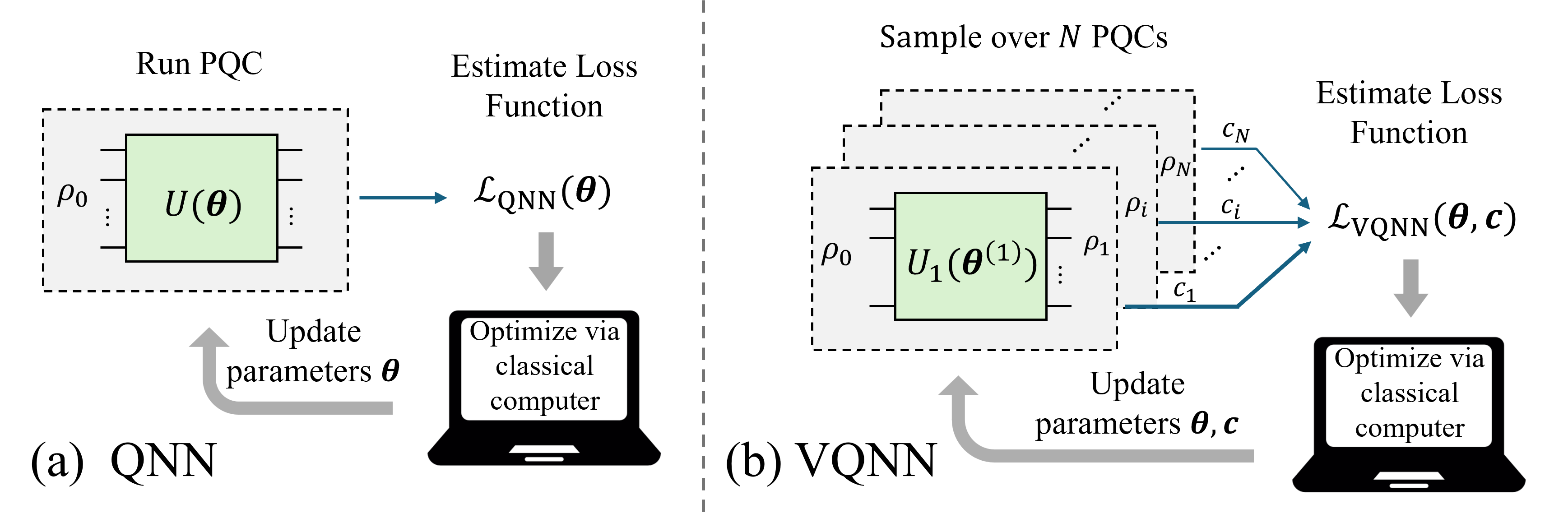}
    \caption{{\bf (a) Training of a quantum neural network.}  The network is designed to implement a parametrized quantum circuit (PQC), depending on a set of parameters $\boldsymbol{\theta}$. In the noiseless case, the PQC is represented by a unitary operator $U(\boldsymbol{\theta})$, which is applied to a quantum system, initially in the state $\rho_0$. Then, a measurement is performed, and the statistics of the outcomes is used to  estimate the value of a loss function $\cL_{\rm QNN}(\boldsymbol{\theta})$. Finally, the classical optimization method is used to update the parameters in the PQC to minimize the loss function. {\bf (b) Training  of a virtual quantum neural network.} $N$ parametrized quantum circuits (PQCs) are randomly sampled and applied to the input state $\rho_0$, producing output states $\rho_1, \dots,  \rho_N$.  Quantum measurements performed on the output states are then used to estimate the value of the loss function $\cL_{\rm VQNN}(\boldsymbol{\theta,c})$, which depends on the parameters of the sampled circuits, and on a set of coefficients $c_1,\dots,  c_N$, used to aggregate the data from the $N$ different branches.  The value of the loss function  is then fed into a  classical optimization algorithm  used to update the value of the parameters in the PQCs.   }
    \label{fig:workflow}
\end{figure*}

We call our extended QNN model  a \textit{virtual} QNN (VQNN) and illustrate its quantum-classical optimization loop in Fig. \ref{fig:workflow}.  Compared with conventional QNNs, VQNNs framework offers two key advantages: enhanced expressivity and improved robustness against quantum noise.   We  demonstrate these features in  three representative quantum machine learning tasks, including both supervised and unsupervised learning. First, we consider the problem of quantum error mitigation, introducing a technique of   variational quantum error mitigation (VQEM).   Numerical results show that the mitigation strategies obtained via VQNNs are near-optimal when implemented with noiseless parameterized quantum circuits, and remain effective even in the presence of circuit-level noise. Second, we develop a binary classification algorithm, which we call  the virtual quantum classifier (VQC). For both training and testing data sets, VQC consistently achieves higher prediction accuracy than conventional QNN-based classifiers. Finally, we apply the VQNN framework to  the problem of ground-state energy estimation,  proposing a virtual variational quantum eigensolver (VVQE). With noisy circuits, VVQE yields significantly more accurate estimates of the ground-state energy compared with standard QNN approaches. Taken together, these results establish VQNNs as a flexible paradigm for quantum machine learning, which enhances  expressivity and reduces sensitivity to noise, thereby advancing the prospects for future practical applications.

The rest of the paper is structured as follows: in Section~\ref{sec:VQNN}, we introduce the VQNN model and discuss its general features.   The applications are then presented in Section~\ref{sec:applications}, where we apply the VQNN model to two supervised learning  tasks, namely  quantum noise mitigation (Subection~\ref{sec:error mitigation}) and quantum classification (Subsection~\ref{sec:classification}), and to an unsupervised learning task, namely the estimation of the ground state energy of a binuclear molecule (Subsection~\ref{sec:VQE}).   Finally, Section \ref{sec:discussion}  provides a discussion of our findings and draws  the conclusions.

\section{Virtual Quantum Neural Networks}\label{sec:VQNN}

\subsection{Standard QNNs}

Before introducing our extended notion, here we briefly summarize the standard notion of QNN~\cite{cerezo2021variational}, illustrated in Fig.~\ref{fig:workflow} (a).  A standard QNN is physically implemented by a parametrized quantum circuit, acting on a given quantum system, typically consisting of a finite number of qubits.    In the noiseless case, the  circuit is  described by a parametrized unitary operator $U(\boldsymbol{\theta})$, acting on the Hilbert space $\cal H$ of the given system and depending on a set of parameters $\boldsymbol{\theta}$.    In the noisy case, the circuit is described by a parametrized quantum channel, that is, a completely positive, trace-preserving linear map $\cC (\boldsymbol{\theta})$ mapping linear operators on the  Hilbert space  $\cal H$ into linear operators on a (possibly different) Hilbert space $\cal K$, and again, depending on a set of parameters $\boldsymbol{\theta}$.  In the following, we will denote by $L({\cal H})$  ($L({\cal K})$)
 the set of linear operators from the Hilbert space $\cal H$ ($\cal K$) to itself.

The training of the QNN, illustrated in Fig. \ref{fig:workflow} (a), is aimed at minimizing the value of a loss function.  In most cases, the loss function is a function of the expectation value of a given observable. In the noiseless case, this type of loss function can be written as    \begin{equation}
\label{eq:cost_funcQNNnoiseless}
\cL_{\rm QNN}(\boldsymbol{\theta}) := f\left(\tr\left[O\, U(\boldsymbol{\theta} )\rho_0 U^\dagger(\boldsymbol{\theta})\right]\right) \, ,
\end{equation}
where $\rho_0 \in L({\cal H})$ is a density matrix ($\tr[\rho_0]=1$, and $\<\psi|  \rho_0|\psi\>  \ge 0 \,, \forall |\psi\> \in \cal H$), representing a state of the quantum system, $O \in L({\cal K})$ is a Hermitian operator ($O^\dag =  O$), representing an observable, and $f:  {\mathbb R} \to {\mathbb R}$ is a  real-valued function depending the task at hand. A basic requirement for training is that the loss function is lower bounded, so that a finite minimum exists, in the same way as in the noiseless  case.  Other requirements, such as differentiability, may be further imposed depending on the minimization algorithm used to minimize the loss function.

In the noisy case, the loss function takes the more general form
\begin{equation}\label{eq:cost_funcQNN}
    \cL_{\rm QNN}(\boldsymbol{\theta}) =f\left(\tr\left[O\, \cC(\boldsymbol{\theta} )(\rho_0)  \right]\right) \, ,
\end{equation}
where $\rho_0 \in L({\cal H})$ is a quantum state of the input system,  $O_k \in L({\cal K})$ is an observable on the output system, and  $f:  {\mathbb R} \to {\mathbb R}$ is a real-valued function associated to the task at hand.

Both in the noiseless and noisy case, the value of the loss function can be directly estimated by initializing the input system in the state $\rho_0$, running the circuit,  and performing a measurement of the observable $O$ on the final state. The estimated value of the loss function is then fed into a classical optimizer, which updates the parameters $\boldsymbol{\theta}$, searching for the value that minimizes the loss, in the same way as in the noiseless case.

\subsection{Noiseless VQNNs}

We are now ready to introduce the framework of VQNNs.
We will start from its basic noiseless version, illustrated in Fig.~\ref{fig:workflow} (b). In the noiseless case, a VQNN is specified by $N$ real coefficients $c_1, \dots, c_N$, and by $N$  unitary operators $U_1(\boldsymbol{\theta}^{(1)}), \dots ,  U_N(\boldsymbol{\theta}^{(N)})$, where $\boldsymbol{\theta}^{(i)}$ are the parameters of the $i$-th circuit, for $i\in  \{1,\dots, N\}$.   Each of these unitary operators represents a parametrized quantum circuit (PQC) acting on a given quantum system with Hilbert space $\cal H$.

The loss function is defined in terms of an observable $O$, mathematically described by a Hermitian operator $O  \in L ({\cal H})$: explicitly, the loss function is
\begin{equation}\label{eq:general_cost_func_noiseless}
    \cL_{\rm VQNN}(\boldsymbol{\theta,c}) :=f\left( \sum_{i=1}^N c_i  \tr\left[O\, U_i(\boldsymbol{\theta}^{(i)})\rho_0 U_i^\dagger(\boldsymbol{\theta}^{(i)})\right]\right) \, ,
\end{equation}
where we used the notations $\boldsymbol{c}  :  = (c_1, \dots, c_N)$ and  $\boldsymbol{\theta}  :  = \left(\boldsymbol{\theta}^{(1)}, \dots, \boldsymbol{\theta}^{(N)}\right)$, for the coefficients and the parameters, respectively. In the above equation, $\rho_0 \in L({\cal H})$ is a density matrix, representing a state of the quantum system, while   the function $f: \mathbb{R} \to \mathbb{R}$
 depends on the task at hand, as in the case of standard QNNs.

The loss function Eq.~\eqref{eq:general_cost_func_noiseless} can be empirically estimated from measurement data.  More precisely, measurement data can be used to obtain an empirical estimate of the loss function's argument
\begin{align}\label{lossfunctionargument}
x  (\boldsymbol{\theta},\boldsymbol{c}
):  = \sum_{i=1}^N c_i  \tr\left[O\, U_i(\boldsymbol{\theta}^{(i)})\rho_0 U_i^\dagger(\boldsymbol{\theta}^{(i)})\right]\, .
\end{align}
The estimation procedure is as follows.     First, the quantum system is initialized in the state $\rho_0$.  Then, it undergoes the unitary gate $U_i(\boldsymbol{\theta}^{(i)})$    with probability $p_i:  = |c_i|/\|\boldsymbol{c}\|_1$, with
\begin{align}\label{gamma}
\| \boldsymbol{c}\|_1:  = \sum_{j=1}^N |c_j|  \, .
\end{align}
Due to the action of the gate, the system  ends up  in the state $\rho_i (\boldsymbol {\theta}) :=  U_i(\boldsymbol{\theta}^{(i)})  \rho_0  \, U_i(\boldsymbol{\theta}^{(i)})^\dag$.
Finally, the expectation value of the observable $O$ is estimated, {\it e.g.} by performing a von Neumann measurement on the eigenstates of the operator $O$, which yields an outcome $o_k $ with probability $p_k^{(i)}   :  = \tr[  P_k\, \rho_i (\boldsymbol {\theta}) ]$ where $o_k$ is one of the eigenvalues of $O$ and  $P_k$ is the projector on the corresponding eigenspace.

The above procedure is repeated for $S$ times, every time sampling one of the unitary gates $(U_1(\boldsymbol{\theta}^{(1)}), \dots,  U_N(\boldsymbol{\theta}^{(N)}))$ according to the probability distribution $(p_1,\dots, p_N)$.  After the $S$ rounds of sampling have been concluded, the measurement outcomes are used  to compute the  linear combination
\begin{equation}\label{eq:exp_estimation}
    \zeta_S=\frac{\|\boldsymbol{c}\|_1}{S} \sum_{s=1}^S \frac{c_{i(s)}}{|c_{i  (s)}|}   \,  o_{k(s)},
\end{equation}
where the index $s\in\{1,\dots,  S\}$ labels the round, $k(s)$ is the  label  of the  measurement outcome obtained in the $s$-th round, and $i(s)$ is the label of the unitary gate  chosen in the $s$-th round.

For large $S$, the law of large numbers implies that the quantity $\zeta_S$ in Eq.~\eqref{eq:exp_estimation} converges with high probability to $x  (\boldsymbol{\theta},\boldsymbol{c})$, the argument of the loss function.  Quantitatively,  Hoeffding’s inequality~\cite{hoeffding1963probability} guarantees that $\zeta_S$  has probability at least  $1-\delta$ to deviate at most $\varepsilon$ from $x  (\boldsymbol{\theta},\boldsymbol{c})$, provided that the number of sampling rounds is at least
\begin{equation}\label{eq:hoeffding}
S =  2\frac{ \|  \boldsymbol{c}\|_1^2 \,\|O\|_\infty^2}{\varepsilon^2}\ln\left(\frac{2}{\delta}\right) \,,
\end{equation}
$\|O\|_\infty:=\max_k |o_k|$ is the operator norm of the observable $O$. Note that, for a fixed  observable $O$ and error parameters $\epsilon$ and $\delta$,  the sampling time $S$ depends only on $\|\boldsymbol{c}\|_1$.  For this reason, $\|\boldsymbol{c}\|_1$  is  called the~\textit{sampling overhead}~\cite{jiang2021physical,regula2021operational,zhao2025power}.   Finally, the value of $\zeta_S$ is plugged into the function $f$, thereby providing an estimate of the loss function $\cL(\boldsymbol{\theta, c})=f(x (\boldsymbol{\theta}, \boldsymbol{c}))$.

After estimating the loss function $\cL(\boldsymbol{\theta,c})$, classical optimization methods~\cite{mitarai2018quantum,stokes2020quantum,parrish2019jacobi, nakanishi2020sequential,ostaszewski2021structure} are used to update the parameters of the PQCs in order to minimize the loss function.  This   classical-quantum optimization loop is iterated until the loss function converges to a minimum.  In this work, the search for the minimum has been carried out using   \textit{Adam}~\cite{kingma2014adam}, a popular optimization algorithm based on stochastic gradient descent.

It is worth mentioning that the training  of standard QNNs may be subject to the problem of barren plateaus~\cite{larocca2025barren}, arising when the gradients of the loss function  vanish exponentially with the size of the system.  A similar issue can  affect  also VQNNs; however,  the additional flexibility of the VQNN framework does not generally come at the price of aggravating the problem of  barren plateaus, as discussed in  Appendix~\ref{app:BP}.

\subsection{Noisy VQNNs}

In general, NISQ circuits are noisy, and instead of implementing unitary operators, they implement quantum channels, mathematically described by CPTP maps.  The formulation of the VQNN model in the noisy  setting is a straightforward modification of the noiseless formulation: a noisy VQNN is specified by $N$ real coefficients $c_1, \dots, c_N$, and by $N$ parametrized CPTP maps $\cC_1(\boldsymbol{\theta}^{(1)}),  \dots, \cC_N(\boldsymbol{\theta}^{(N)})$, each of which transforms an input system, with Hilbert space $\cal H$,  into an output system, with  Hilbert space $\cal K$.

The training consists in the minimization of the
loss function
\begin{equation}\label{eq:general_cost_func}
    \cL_{\rm VQNN}(\boldsymbol{\theta,c}) := f\left(\sum_{i=1}^N c_i\tr\left[O\, \cC_i(\boldsymbol{\theta}^{(i)}) (\rho_0)  \right]\right) \, ,
\end{equation}
where $\rho_0 \in L({\cal H})$ is a density matrix on the input system,  $O \in L({\cal K})$ is an  observable on the output system, and  and $f:  {\mathbb R} \to {\mathbb R}$ is a real-valued function depending on the problem at hand.

The loss function can be empirically estimated by preparing the input system in the initial state $\rho_0$,  applying the channel $\cC_i ( \boldsymbol{\theta}^{(i)})$ with probability $p_i  =  |c_i|/\|  \boldsymbol{c}\|_1$, measuring the observable $O$, and postprocessing the outcomes, in the same way as it was described in the noiseless case.

\subsection{Comparison to standard QNNs}

A standard QNN can be seen as the special case of a VQNN with a single branch  ($N=1$), corresponding to a single parametric family of quantum channels $\cC_1 (\boldsymbol{\theta}^{(1)} )$,  with the corresponding coefficient set to the default value $c_1=  1$. Since there is only one branch, here we will drop the branch index, writing $\boldsymbol{\theta}$ and $\cC (\boldsymbol{\theta} )$  in place of $\boldsymbol{\theta}^{(1)}$ and    $\cC_1 (\boldsymbol{\theta}^{(1)} )$, respectively.    In this case, the loss function (\ref{eq:cost_funcQNN}) is simply equal to the function $f$ applied to the expectation value of the observable $O$ on the output state $\cC (\boldsymbol{\theta})  (\rho_0)$.

For a general VQNN, instead, the loss function Eq.~\eqref{eq:general_cost_func} can be rewritten as
\begin{equation}\label{eq:general_cost_funcHP}
    \cL_{\rm VQNN}(\boldsymbol{\theta,c}) :=f\left( \tr\left[O\, \cM  (\boldsymbol{\theta,c}) (\rho_0)  \right]\right) \, ,
    \end{equation}
with
\begin{align}\label{eq:HP_map}
\cM  (\boldsymbol{\theta,c})   : =   \sum_{i=1}^N c_i\,  \cC_i(\boldsymbol{\theta}^{(i)})  \,.    \end{align}

The key difference between a standard QNN and a VQNN can be seen by comparing the loss functions in  Eq.~\eqref{eq:cost_funcQNN} and Eq.~\eqref{eq:general_cost_funcHP}, respectively. For a standard QNN, the loss function is minimized over a parametric family of quantum channels,  mathematically described by CPTP maps.  For a virtual QNN, instead, the loss function is minimized over a larger parametric family,  containing all  linear combinations of the  CPTP maps associated to the $N$ branches.  In general, such linear combinations are Hermitian-preserving (HP) maps, that is, linear maps that transform Hermitian operators into Hermitian operators. HP maps are also known as \textit{virtual quantum operations}, and  have applications in several quantum tasks, including entanglement detection~\cite{peres1996separability, horodecki2001separability, guhne2009entanglement}, reduced dynamics of open quantum systems~\cite{pechukas1994reduced,modi2012operational,schmid2019initial}, and quantum error mitigation~\cite{temme2017error,takagi2021optimal,takagi2022fundamental,jiang2021physical,zhao2023information,zhao2024retrieving}.

In summary, the loss function   for a VQNN  is equal to a function $f$ over the expectation value of the observable $O$ on the output of the virtual quantum operation $\cM (\boldsymbol{\theta,c})$, applied to the initial state $\rho_0$. By extending the optimization space from a parametric family of CPTP maps to a larger parametric family of HP maps,   VQNNs offer increased expressivity and performance by leveraging a larger optimization space. The price to pay for these enhancements is an increased sampling time, according to the sampling overhead $\gamma$, and an increased classical computational effort, required for  minimizing the cost function over a larger set of parameters.  In other words, the VQNNs framework can be seen as a way to overcome some of the limitations of the  available quantum hardware by investing additional resources into sampling and classical computation.   It is also worth noting  that the $N$ branches of the VQNN can be executed in parallel, meaning that the VQNN framework is naturally suitable for the distributed computing paradigm.

\subsection{Further extensions}

A further extension of our model is obtained by sampling over quantum circuits that include measurement devices.   Mathematically, a quantum circuit with $M$ possible outcomes  is represented by a quantum instrument, that is, an indexed set  of completely positive maps $(\cC_j)_{j=1}^M$, with the map $\cC_j$ corresponding to  the $j$-th outcome. The normalization condition for a quantum instrument is that the sum $\sum_{j=1}^M  \cC_j$ is a CPTP map.

In the most general case, a VQNN can be implemented by sampling over $N$ quantum instruments, each of which is parametrized by a set of parameters.  The interest of this generalization is that the outcomes of the instruments can be used in the postprocessing stage, thereby computing linear combinations of the expectation values conditional on different measurement outcomes. This outcome-dependent postprocessing allows one to compute linear combinations  of the form
\begin{align}\label{lossfunctionargumentinstrument}
x  (\boldsymbol{\theta},\boldsymbol{c}
):  = \sum_{i=1}^N \sum_{j=1}^M  c_{ij}  \tr\left[O\, \cC_{ij}(\boldsymbol{\theta}^{(i)})(\rho_0) \right]\, ,
\end{align}
where    $\left(\cC_{ij} (\boldsymbol{\theta}_i) \right)_{j=1}^M$ denotes the $i$-th quantum instrument, $(c_{ij})_{j=1}^M$ are the corresponding coefficients, $\boldsymbol{\theta}  :=  (\boldsymbol{\theta}_1, \dots, \boldsymbol{\theta}_N)$ are the  parameters of the $N$ circuits, and $\boldsymbol{c}:= (c_{ij})_{1\le i\le N,\,  1\le j\le M}$ is the matrix of all coefficients.
The training of the VQNN is then designed to minimize the value of a loss function  of the form $\cL (\boldsymbol{\theta},\boldsymbol{c}) :  =f(x  (\boldsymbol{\theta},\boldsymbol{c})
)$ where $f: \mathbb{R} \to \mathbb{R}$ is a function depending on the task at hand.

Computing the  linear combination (\ref{lossfunctionargumentinstrument}) by classical postprocessing is equivalent to simulating the linear map
\begin{align}\label{eq:HP_mapprobabilistic}
\cM  (\boldsymbol{\theta,c})   : =   \sum_{i=1}^N \sum_{j=1}^M c_{ij}\,  \cC_{ij}(\boldsymbol{\theta}^{(i)})  \,,    \end{align}
This expression is more general than the expression in  Eq. (\ref{eq:HP_map}), which consisted in a linear combination of \textit{trace-preserving} maps, and, as such, rescaled the trace of every matrix $\rho$ by a fixed amount $\lambda  :  = \sum_{i=1}^N  c_i$.  In contrast, Eq. (\ref{eq:HP_mapprobabilistic})  is a linear combination of \textit{trace non-increasing} maps, and, as such, it can be an arbitrary HP map, without any further constraint on the trace.   Hence, the extension to quantum instruments provides a further enlargement of the optimization space.

Finally, it is worth mentioning that a special case of the VQNN model is to use quantum circuits that act non-trivially only on a subsystem of the total quantum system.   In this case, the quantum circuits are used to compute the expectation value
\begin{align}\label{lossfunctionargumentinstrumentallstuff}
x  (\boldsymbol{\theta},\boldsymbol{c}
):  = \sum_{i=1}^N \sum_{j=1}^M  c_{ij}  \tr\left[O\, \left(\cC_{ij}(\boldsymbol{\theta}^{(i)}) \otimes \cI_R\right) \,(\rho_0) \right]\, ,
\end{align}
where the input quantum system is divided into two subsystems $Q$ and $R$,   with  the identity channel $\cI_R$ acting on subsystem $R$, and  the sampled quantum instruments acting on subsystem $Q$, transforming it into a new system $Q'$.   By choosing $\rho_0$ and $O$ to be an arbitrary state of system $QR$ and an arbitrary observable on system $Q'R$, this scheme allows one to realize every possible test on the HP map $\cM  (\boldsymbol{\theta,c})$ \cite{bai2018test}, thereby allowing for a broader set of tasks compared to the basic formulation without the reference system.

\section{Applications}\label{sec:applications}

In this section  we apply the VQNN framework to three different problems, including two supervised learning tasks (quantum noise mitigation and quantum classification) and an unsupervised learning task (ground-state energy estimation). For simplicity,  we will consider VQNNs realized by sampling over  $N=2$ quantum circuits, which is already sufficient to achieve  non-trivial Hermitian-preserving maps~\cite{jiang2021physical, zhao2023information}.
%In this case,  the coefficients  of the two branches are in one-to-one correspondence with the sampling overhead $\gamma$,  via the relations $c_1 = \frac{1+\gamma}{2}$ and $c_2 = \frac{1-\gamma}{2}$.
In all these applications, we  will use the \textit{QUAIRKIT} platform~\cite{quairkit} to conduct numerical experiments to demonstrate the advantage of VQNN on training quantum neural networks with noisy quantum devices.

\subsection{Mitigating  unknown quantum noise}\label{sec:error mitigation}

Quantum error mitigation~\cite{jiang2021physical, zhao2023information, zhao2024retrieving} is a promising approach to estimate the expectation values of quantum observables in the presence of noise. The existing protocols, however, require the action of the noise to be known, either \textit{a priori} or via quantum process characterization. Here we use the VQNN framework to develop a quantum error mitigation algorithm that can be used in the presence of unknown, uncharacterized noise.

\subsubsection{Standard quantum error mitigation}

Estimating the expectation values of quantum observables is a key step in several quantum algorithms~\cite{cerezo2021variational, nielsen2010quantum,peruzzo2014variational}. In the ideal scenario, the expectation value $\tr[O\rho]$ of a given observable $O$ on a given quantum state $\rho$ is estimated  by directly performing measurements on the state $\rho$.    In practice, however, quantum noise is inevitable, and one has access to a noisy  version of the quantum state $\rho$, denoted by $\cN(\rho)$ where $\cN:  L ({\cal H}_A) \to L({\cal H}_B)$ is a  CPTP map. Then, the problem of quantum error mitigation is to estimate the expectation value  $\tr[O\rho]$ by performing measurements on the noisy state $\cN(\rho)$.

The key idea of quantum error mitigation is to invert the action of the noise, by applying the inverse map $\cN^{-1}$.    This  map is generally Hermitian-preserving and trace-preserving (HPTP) map, but not CPTP.  As such,  it cannot be implemented directly on a quantum device, but it can be simulated by sampling and postprocessing.  To simulate the map $\cN^{-1}$, one  decomposes it into a linear combination of CPTP maps
\begin{align}
\cN^{-1} = \sum_{i=1}^N c_i\cC_i \,,
\end{align}
where $c_i$ are real coefficients and $\cC_i$ are CPTP maps~\cite{jiang2021physical, takagi2021optimal, takagi2022fundamental}. Then, the map $\cN^{-1}$ is simulated by sampling over the CPTP maps  $\cC_1,\dots, C_N$, and by postprocessing the expectation values to compute the linear combination.

The optimal error mitigation protocol corresponds to the decomposition that minimizes the sampling overhead, and can be found by solving the following minimization problem:
\begin{align}\label{eq:original_problem}
    \min_{c_i, \cC_i} \Big\{ \|  \boldsymbol{c}\|_1\,  \Big| & \sum_i c_i\cC_i = \cN^{-1}, c_i\in\mathbb{R}, \cC_i\in \text{CPTP}\Big\} \, ,
\end{align}
where $\text{CPTP}$ denotes the set of all CPTP maps from $L({\cal H}_B)$ to $L({\cal H}_A)$. It is important to note that the optimization problem in Eq.~\eqref{eq:original_problem} is defined in terms of a given CPTP map $\cN$, which specifies the action of the noise.  In the following, we will extend the framework of quantum error mitigation to the scenario where the map $\cN$ is initially unknown, corresponding to an uncharacterized quantum noise.

\subsubsection{Variational quantum error mitigation}

We now use the VQNN framework to develop  variational algorithms for mitigation  of an initially unknown noise process. The first step to construct such a protocol is to rephrase  the optimization problem  in Eq.~(\ref{eq:original_problem}) in a way that facilitates variational optimization. We start by observing  that the error mitigation condition    $\cN^{-1}=\sum_ic_i\cC_i$ is equivalent to  $\cI  =   \cN^{-1} \circ \cN =  \sum_ic_i~ \cC_i\circ\cN$, which in turn is equivalent to  \begin{align}\label{bbb}
\Phi^+   =    \sum_i  \, c_i\,  ( \cI_A\otimes \cC_i \circ \cN)  (\Phi^+)  \, ,
\end{align}\
where $\Phi^+   :=  |\Phi^+\>\<\Phi^+|$ is the projector onto the Bell state $|\Phi^+\>  =  \sum_{i=1}^{d_A}  |i\>\otimes |i\>/\sqrt{d_A}$, and $d_A$ is the dimension of system's $A$ Hilbert space. The derivation of Eq.~\eqref{bbb} is given in Appendix~\ref{app:derivatino_of_eq}.
%and $A'$ and $A''$ are quantum systems with $d_A$-dimensional Hilbert space.

Eq. (\ref{bbb}) guarantees a perfect inversion of the noise, corresponding to an ideal error mitigation.  In practice, however, a perfect inversion may  not possible, either because the map $\cN$ is not  invertible, or because the operations  accessible on the given quantum hardware do not include the channels $(\cC_i)_{i=1}^N$  needed to invert the noise. In this setting, Eq. (\ref{bbb}) can only hold approximately, and the goal is to find the best approximation achievable with the existing hardware.   To assess the quality of an approximation in Eq. (\ref{bbb}), one can use a measure of distance between the ideal target state $\Phi^+$ and the  operator
\begin{align}\label{sigma}\sigma:    = \sum_i  \, c_i\, \sigma_i  \, , \qquad \sigma_i : = ( \cI_A\otimes \cC_i \circ \cN)  (\Phi^+)\, .
\end{align}   In general, this operator may not be positive, and therefore may not be  a quantum state.   We can regard it as  a ``virtual quantum state,'' which can be simulated by random sampling and classical data processing.

Many choices of distance measures are possible, including for example the fidelity and the diamond norm. For variational optimization, however, it is important to choose a distance measure that can be easily estimated from experimental data, so that one can feed the estimate of the distance into the classical algorithm that optimizes the parameters of the channels $(\cC_i)_{i=1}^N$.
 For this purpose, a suitable choice is provided by the Hilbert-Schmidt (HS) distance, used in earlier works on quantum process tomography  \cite{scott2008optimizing,bisio2009optimal1,bisio2009optimal2} and, more recently, in quantum state learning and purification~\cite{chen2021variational}.   Mathematically, the HS distance between the operators $\Phi^+$ and $\sigma$ is
\begin{align}
\nonumber
    \left\|\Phi^+-\sigma\right\|_2^2
   &:= \tr[(\Phi^+-\sigma)^2] \\
   \nonumber &=\tr[(\Phi^+)^2] + \tr[\sigma^2] -2\tr[\Phi^+\sigma]\nonumber\\
    &= 1 + \sum_{i,j=1}^N c_ic_j\tr[\sigma_i \sigma_j] - 2\sum_{i=1}^N c_i \tr[\Phi^+ \sigma_i]\, .
\end{align}
Crucially, the overlaps $\tr[\sigma_i \sigma_j]$ and $\tr[\Phi^+ \sigma_i]$ can be empirically estimated from quantum measurements.  Specifically, for the first overlap $\tr[\sigma_i \sigma_j]$, one can be estimated by applying the swap test~\cite{barenco1997stabilization,buhrman2001quantum, nielsen2010quantum} on the states $\sigma_i$ and $\sigma_j$ for arbitrary $i,j\in\{1,\cdots, N \}$. Moreover, if the control SWAP operation is unavailable, one can estimate the overlap by local operation and classical operations instead~\cite{anshu2022distributed}. For the other overlap,  $\tr[\Phi^+ \sigma_i]$,  the value can also be estimated from the correlations of local Pauli measurements, using the formula  $\tr[\Phi^+\sigma_i] =( 1+ \tr[(X\otimes X)\, \sigma_i] - \tr[(Y\otimes Y)\,\sigma_i] + \tr[(Z \otimes Z)\, \sigma_i])/4$ (this formula  follows from  the decomposition  of the Bell state as a linear combination of Pauli products, namely  $\Phi^+=(I \otimes I+X \otimes X-Y\otimes Y+Z\otimes Z)/4$). 

\begin{figure}[t]
    \centering
    \includegraphics[width=\linewidth]{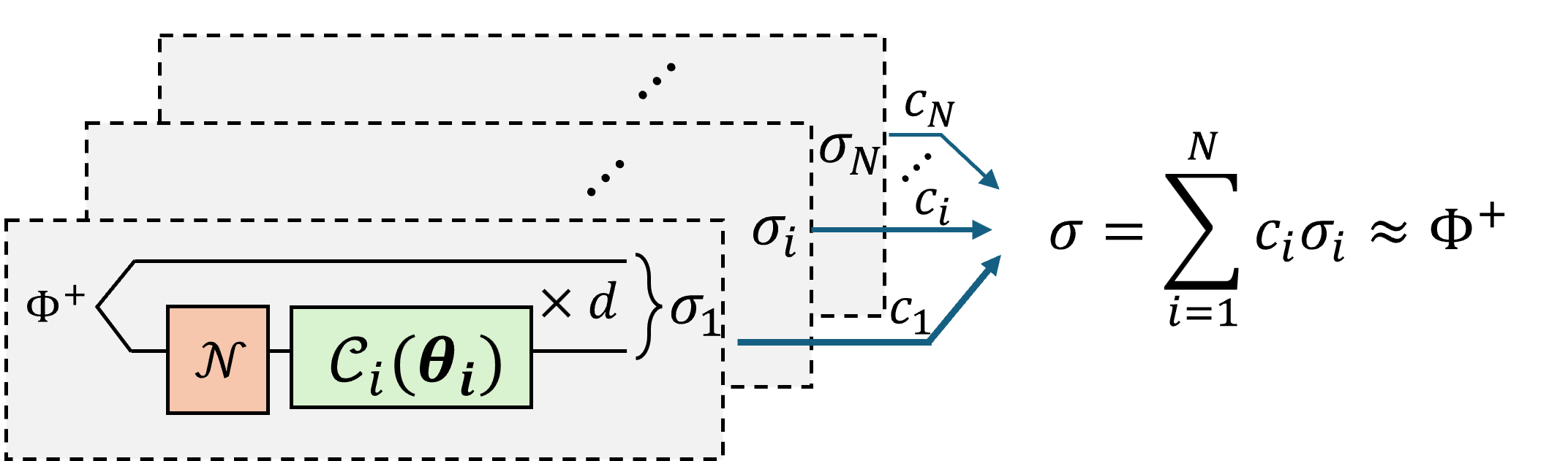}
    \caption{{\bf Variational quantum error mitigation protocol.}  Multiple parametrized quantum channels $\cC_1 (\boldsymbol{\theta_1}), \dots, \cC_N (\boldsymbol{\theta_N})$ are randomly sampled, producing output states $\rho_1,\dots, \rho_N$, respectively. At the postprocessing stage, the output states are combined linearly with  coefficients $(c_i)_{i=1}^N$,  yielding the virtual state $\sigma=\sum_{i=1}^N c_i\rho_i$. The goal of the variational protocol is to find parameters such that the virtual state $\sigma$  is is close to the Bell state $\Phi^+$.}
    \label{fig:error_mitigation_diagram}
\end{figure}

We are now ready to formulate the error mitigation task in the VQNN framework. The basic scenario is illustrated in Fig. \ref{fig:error_mitigation_diagram}.   Here,  we let the channels $(\cC_i)_{i=1}^N$ depend on variational parameters $(\boldsymbol{\theta}_i)_{i=1}^N$, respectively, and train our VQNN to minimize  the  approximation error.   We propose two algorithms, corresponding to two different ways to include the sampling overhead into the optimization.   The first algorithm is based on the minimization of the loss function
\begin{align}\label{eq:VQEM_loss_function_1}
    \cL_{\rm VQEM}^{(1)}(\boldsymbol{\theta}, \boldsymbol{c})=\| \boldsymbol{c} \|_1 + \alpha \, \left\|\Phi^+-\sigma  (\boldsymbol{\theta})\right\|_2^2 \,,
\end{align}
where   $\sigma (\boldsymbol{\theta})  : = \sum_{i}  c_i\, \left(\cI_A \otimes \cN_i(\boldsymbol{\theta}_i) \right) (\Phi^+) $ is the virtual state produced by the VQNN, and $\alpha \ge 0$ is a weight determining the relative importance of the sampling overhead $\|  \boldsymbol{c}\|_1$ and the approximation error $\left\|\Phi^+-\sigma  (\boldsymbol{\theta})\right\|_2$.  In the following, we will refer to $\alpha$
as the \textit{penalty strength}, as it quantifies how strong is the penalty associated to an approximate realization of the error mitigation condition.

The second algorithm aims at directly minimizing the approximation error, and corresponds to the loss function
\begin{align}\label{eq:VQEM_loss_function_2}
    \cL_{\rm VQEM}^{(2)}(\boldsymbol \theta) =  \left\|\Phi^+-\sigma  (\boldsymbol{\theta})\right\|_2^2  \, .
\end{align}
In this case,  the minimization is performed under the constraint that the sampling overhead must not exceed  a given value $\gamma_*$, hereafter called  the \textit{sampling overhead budget}.  Mathematically, this constraint amounts to the fact that  the coefficients $\boldsymbol{c}$ are restricted  to  be in the $N$-dimensional hypercube  defined by the inequality $\sum_{i=1}^N  \, |c_i |  \le \gamma_*$.
Explicit descriptions of the two algorithms in pseudo-code are provided in Appendix ~\ref{app:VQEM}.

In the following we apply the general framework of  variational quantum error mitigation to two special cases, corresponding the mitigation of single-qubit noisy channels using  two different forms of the  parametrized channels $\left(\cC_i  (\boldsymbol{\theta}_i)\right)_{i=1}^N$. In the first case, we take the parametrized channels to be a subset of single-qubit channels generated by bounded-depth unitary gates acting on a limited number of qubits. This scenario models the situation in which a small-size, but otherwise ideal quantum circuit is used to mitigate an initially unknown error.  In the second case we consider the realistic case in which also the realization of the quantum circuit is subject to noise.  We take these two cases as an opportunity to illustrate our two variational algorithms, corresponding to the two loss functions (\ref{eq:VQEM_loss_function_1}) and (\ref{eq:VQEM_loss_function_2}), respectively.

\subsubsection{VQEM with bounded-size unitary circuits}\label{sec:mitigation_with_ideal_circuit}

Here we consider the problem of mitigating an initially unknown    single-qubit noisy channel $\cN$. To this purpose, here we will use our first VQEM algorithm, designed to minimize the loss function (\ref{eq:VQEM_loss_function_1}) with a given penalty strength $\alpha$. To test the performance of our algorithm,  we pick the noisy channel $\cN$ from  two possible families channels:  depolarizing channels, corresponding to  CPTP maps of the form
\begin{align}
\cD_p(\rho)=(1-p)\rho + p\frac{I_d}{d}\, , \end{align}
(where $d$ is the dimension of the system and $I_d$ is the $d\times d$ identity matrix), and amplitude damping channels, corresponding to  CPTP maps of the form
  \begin{align}  \cA_q  (\rho)  =  A_0 \rho A_0^\dag  +  A_1 \rho  A_1^\dag
  \end{align}
  with  $A_0\equiv \ketbra{0}{0} + \sqrt{1-q}\ketbra{1}{1}$, and $A_1 \equiv\sqrt{q}\ketbra{0}{1}$.
  In the above expressions the parameters $p$ and $q$ range between 0 and 1 and quantifies the noise level.

\begin{figure}[h]
    \centering
    \includegraphics[width=\linewidth]{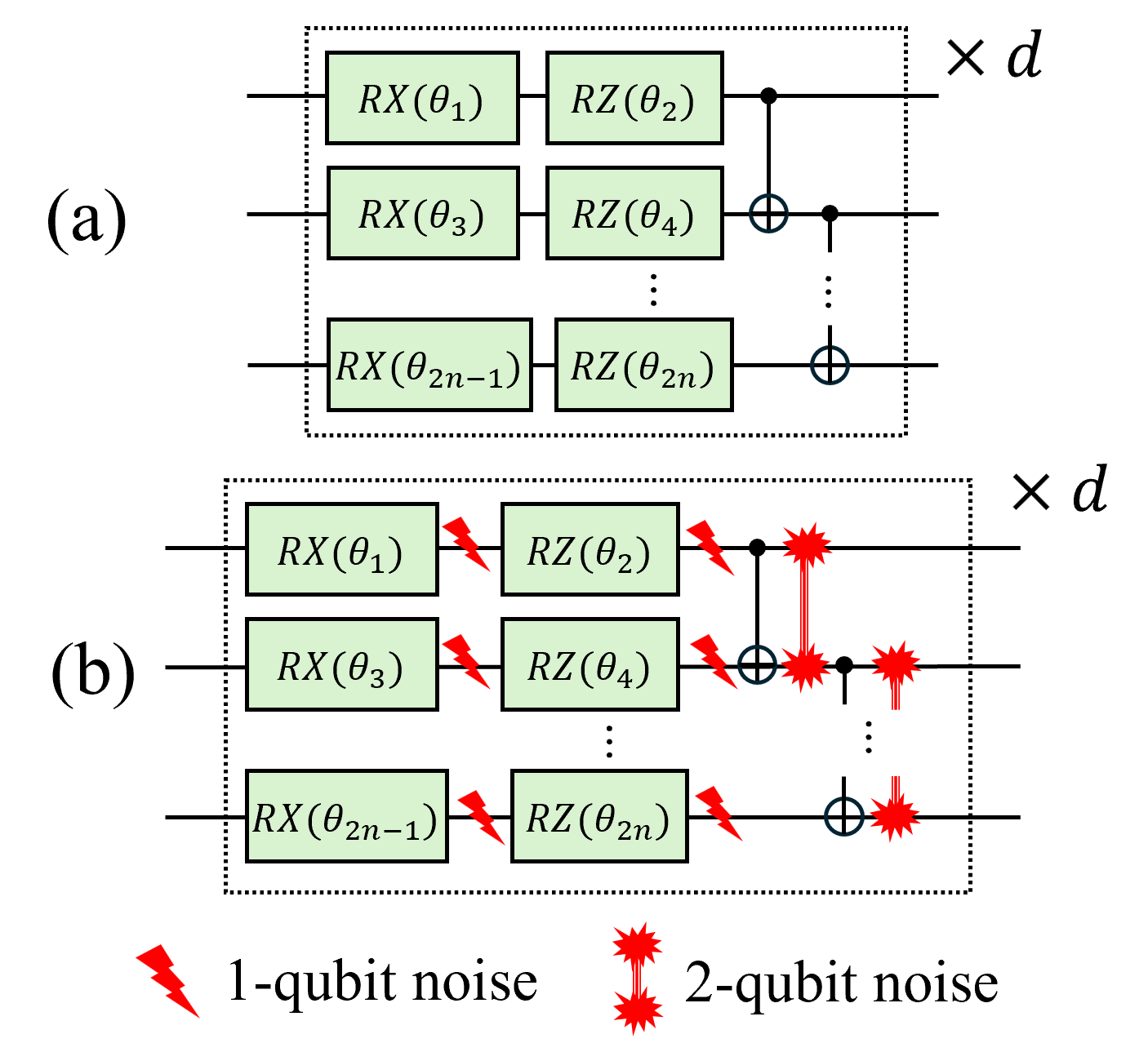}
    \caption{{\bf Variational circuits for error mitigation}. (a) A noiseless ansatz for the PQC is constructed with rotation-X gates, rotation-Z gates and control-NOT gates. (b) A noisy ansatz of PQC is constructed by letting  the single-qubit gates and two-qubit gates be followed by  single-qubit  and two-qubit noise processes, respectively.}
    \label{fig:Ansatz}
\end{figure}

To mitigate the noise, we use parametrized channels obtained by running bounded-size unitary circuits on the target qubit and on a bounded number of auxiliary qubits.  Explicitly, we consider parametrized channels of the form
\begin{align}
\cC_i  (\boldsymbol{\theta_i})   (\rho) =   \tr_{A_1, \dots,  A_{n-1}}   \left[\cU_i  (\boldsymbol{\theta}_i)  \,  (\rho  \otimes |0\>\<0|^{\otimes (n-1)} \right] \, ,
\end{align}
where $A_1, \dots,  A_{n-1}$ are $n-1$ auxiliary qubits, all initialized in the default state $|0\>$, and, for each $i$, $\cU_i  (\boldsymbol{\theta}_i)$ is an $n$-qubit unitary channel a strong entangled layer structure~\cite{zhao2021practical, chen2023near, PhysRevApplied.16.054035,li2024quantum,schuld2020circuit}, corresponding to a unitary gate decomposed into a sequence of $d$ layers, with each layer containing only single qubit rotations about the $X$ and $Z$ axes, and CNOT gates on pairs of neighboring qubits.   For $n=3$,  the strong entangled layer structure is illustrated in   Fig.~\ref{fig:Ansatz} (a).

In our numerical experiments, we set the number of qubits to $n=3$, the number of layers to $d=10$, and the number of parametrized channels to $N=2$.  The first set of experiments consists in the minimization of the loss function in Eq. \eqref{eq:VQEM_loss_function_1}, setting the  penalty strength to $\alpha=1000$.  The results are shown in the TABLE~\ref{tab:sample_overhead}, where we show the values of approximation error and the sampling overhead for the optimal variational circuit.  The approximation error, quantified by the HS norm, appears to be generally small,  showing that the strong entangled layer structure is an effective ansatz for variational error mitigation of depolarizing and amplitude damping channels.

It is also interesting to  observe that sampling overhead of our variational algorithms is  comparable to the absolute minimum of the sampling overhead when the minimization is run over the set of all quantum channels  (see Appendix~\ref{app:SDP} for more details).  In other words,  the circuits identified  by our  VQEM algorithm are nearly optimal in terms of sampling overhead.

\begin{table*}[htbp]
\begin{center}
\begin{tabular}{|c||c|c|c||c|c|c|}
\hline
Noise model & \multicolumn{3}{c||}{Depolarizing $\cD_p$} & \multicolumn{3}{c|}{Amplitude Damping $\cA_q$}\\
\hline
Noise level $p$ & 0.1 & 0.2 & 0.3 & 0.1 & 0.2 & 0.3\\
\hline
HS norm & $9.53\times10^{-7}$ & $1.33\times10^{-6}$& $1.55\times10^{-6}$ & $2.62\times10^{-6}$ & $3.34\times10^{-6}$& $4.53\times10^{-6}$\\
\hline
Sampling overhead by VQEM & 1.1667 & 1.372& 1.644 & 1.222 & 1.500 & 1.857\\
\hline
Minimum sampling overhead & 1.1667 & 1.375 & 1.643 & 1.218 & 1.49467 & 1.849\\
\hline
\end{tabular}
%%%%%%%%%%%%%%%%%%%%%%%%%%%%%%%%%%%%%%%%%%%%%%%%%%%%
% \begin{tabular}{|c|c|c|c|}
% \hline
% Noise model & \multicolumn{3}{|c|}{Amplitude Damping $\cA_q$}\\
% \hline
% Noise level $q$ & 0.1 & 0.2 & 0.3\\
% \hline
% HS norm & $2.62\times10^{-6}$ & $3.34\times10^{-6}$& $4.53\times10^{-6}$\\
% \hline
% Sampling overhead by VQEM& 1.222 & 1.500 & 1.857\\
% \hline
% Minimum sampling overhead & 1.218 & 1.49467& 1.849\\
% \hline
% \end{tabular}
\end{center}
\caption{{\bf Approximation error and sampling overhead of VQEM on depolarizing and amplitude damping noise.}  The approximation error (quantified by the HS norm) and the sampling overhead are shown for different values of the noise level. For comparison, we also include the minimum value of the sampling overhead when the minimization is carried over the set of all single-qubit quantum channels.}
\label{tab:sample_overhead}
\end{table*}

\begin{figure}[h]
    \centering
    \includegraphics[width=\linewidth]{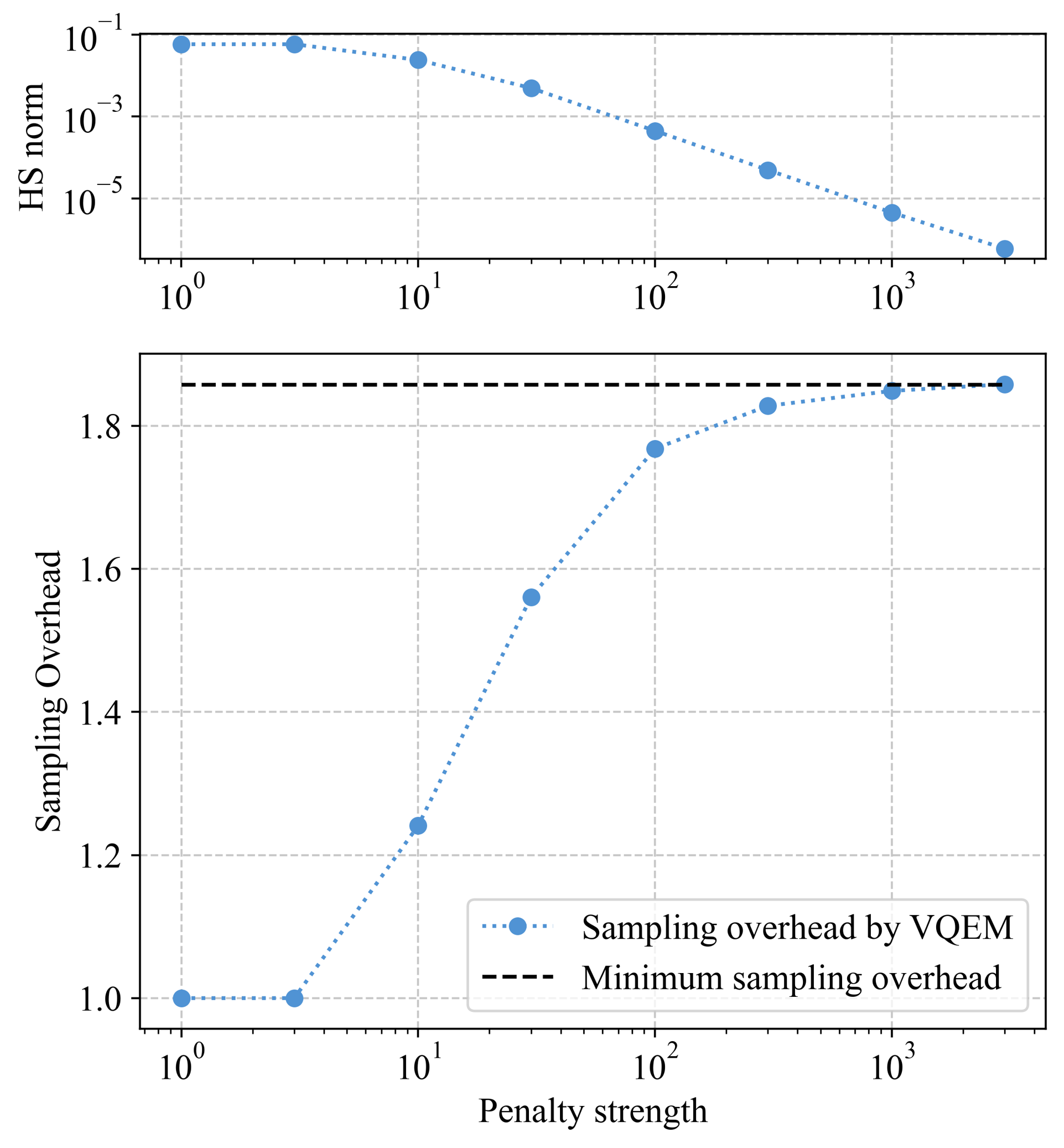}
    \caption{{\bf Dependence of the approximation error and sampling overhead on the penalty strength.} \textit{Top:} relation between the penalty strength  and the approximation error, quantified by the Hilbert-Schmidt  distance between the virtual state $\sigma$ and target Bell state $\Phi^+$.  \textit{Bottom:}  relation between the penalty strength  and the sampling overhead  (blue dotted line).  The minimum sampling overhead over all possible quantum channels is shown by the black dashed line.   }
    \label{fig:penalty}
\end{figure}

It is worth noting that the quality of the approximation decreases for higher values of the noise levels $p$ and $q$. However, this issue can be fixed by tuning the penalty strength $\alpha$, thereby increasing the weight of the HS norm in the loss function.    In Fig.~\ref{fig:penalty}, we analyze the case of  on amplitude damping  with noise level $q=0.3$, showing the approximation error and the sampling overhead as functions of the penalty strength. As anticipated, the approximation error decreases monotonically with the penalty strength.  Notably, also  sampling overhead obtained by running  VQEM  still converges to the minimum sampling overhead over all possible quantum channels.

\subsubsection{VQEM with bounded-size noisy circuits}
We now extend our numerical experiments to the   scenario in which  the parametrized quantum circuits are also subject to noise. As the task, we consider the mitigation of single-qubit depolarizing noise. For the variational optimization, we use our second VQEM algorithm, corresponding to the minimization of the approximation error for a fixed amount of sampling overhead.

To mitigate the noise, we use parametrized channels of the form
\begin{align}
\cC_i  (\boldsymbol{\theta_i})   (\rho) =   \tr_{A_1, \dots,  A_{n-1}}   \left[\widetilde{\cU}_i  (\boldsymbol{\theta}_i)  \,  (\rho  \otimes |0\>\<0|^{\otimes (n-1)} \right] \, ,
\end{align}
where $A_1, \dots,  A_{n-1}$ are $n-1$ auxiliary qubits, all initialized in the default state $|0\>$, and, for each $i$, $\widetilde{\cU}_i  (\boldsymbol{\theta}_i)$ is an $n$-qubit noisy channel, corresponding to a noisy quantum circuit with  a  strong entangled layer structure.  To model noisy circuit, we consider an ideal unitary circuit with strong entangled layer structure, and append a noisy channel to each single-qubit and each two-qubit gate in the circuit, as it was often done in the recent  literature~\cite{takagi2021optimal, scheiber2024reducing,liao2025noise}.
   For $n=3$,  a noisy circuit with  strong entangled layer structure is illustrated in   Fig.~\ref{fig:Ansatz} (b).

In the specific numerical experiments performed here, we set the noise on  single-qubit gates and  two-qubit gates to be depolarizing noise with noise levels  $p=0.01$ and $p=0.02$, respectively.   For the variational circuits, we set the total number of qubits to $n=3$,   the number of layers to $d=5$, and the number of variational channels to $N=2$.   We run our second VQEM algorithm, which minimizes the HS distance with a fixed  sampling overhead  selected in the range of $[1, 2.5]$.  To test the performance of the algorithm we set the channel to be mitigated to a depolarizing channel with noise level  $p=0.2$.

\begin{figure}[h]
    \centering
    \includegraphics[width=\linewidth]{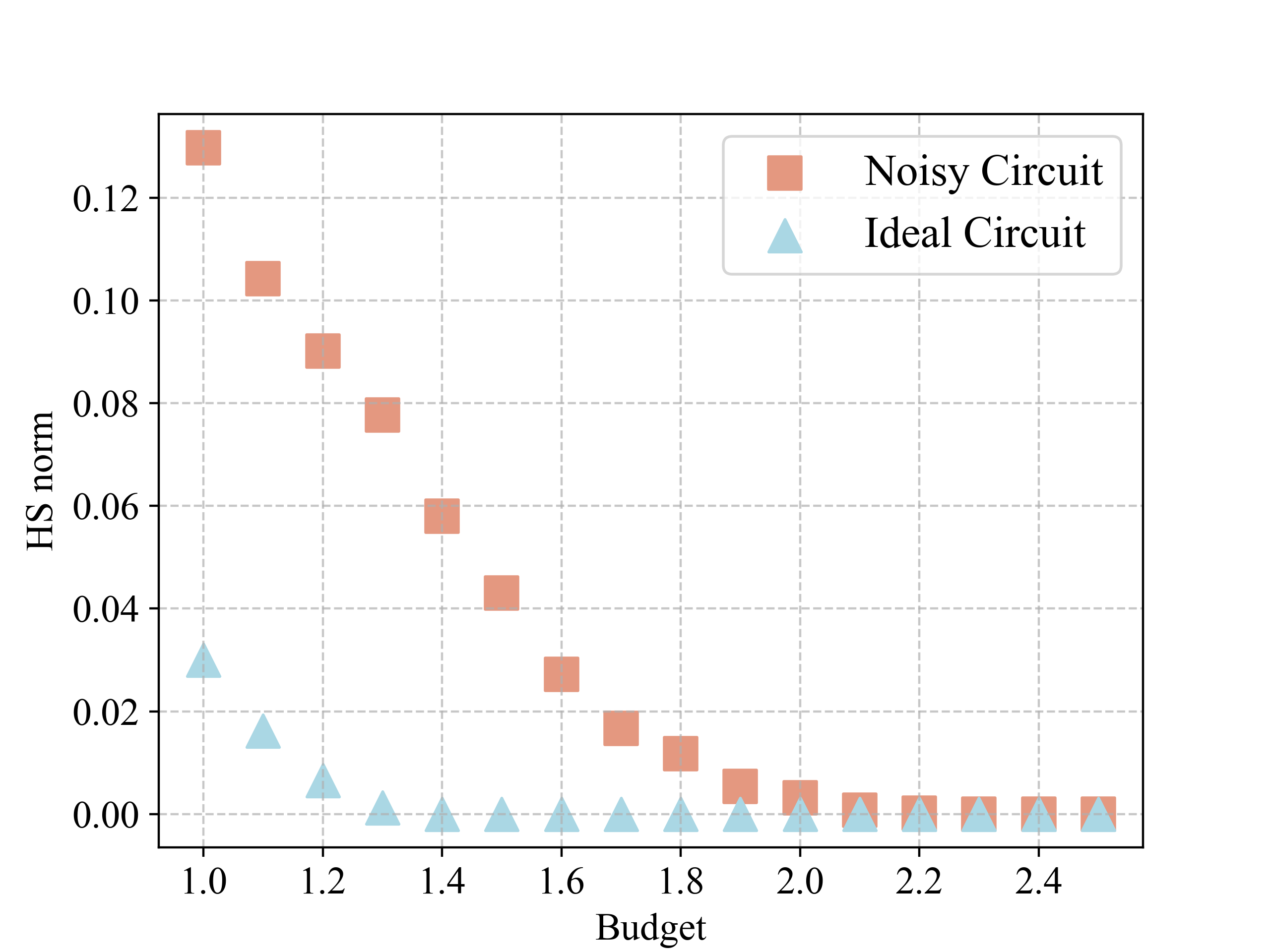}
    \caption{{\bf Approximation error of VQEM on  bounded-size noisy circuits.}  The approximation error, quantified by the HS norm, is shown as a function of the sampling overhead budget in the range $[1,2.5]$.  The orange squares  (light blue squares)  show the results achieved by the VQEM algorithm run on bounded-size noisy (noiseless) circuits. }
    \label{fig:mitigation_with_given_budget}
\end{figure}

The numerical results are shown in Fig.~\ref{fig:mitigation_with_given_budget}, where we compare the optimized HS norm with given sampling overhead budgets by noiseless circuits and noisy circuits. Both HS norms optimized by noisy circuits and noiseless circuits converge to zero with the increase of sampling overhead budget, indicating that the variational form of our circuits still permits a nearly perfect error mitigation.

It is worth observing that, naturally, the HS norm obtained by the noisy circuit is larger than that of the noiseless circuit.   Remarkably, however,  the noisy circuit can nearly reach the same error level of the noiseless one when we allow a larger sampling overhead. Overall, this result shows an important feature of the VQNN model:   random sampling and classical postprocessing can be used to reproduce the performance of an ideal noiseless circuit that is not directly accessible on a realistic quantum hardware.

\begin{figure*}[t]
    \centering
    \includegraphics[width=\linewidth]{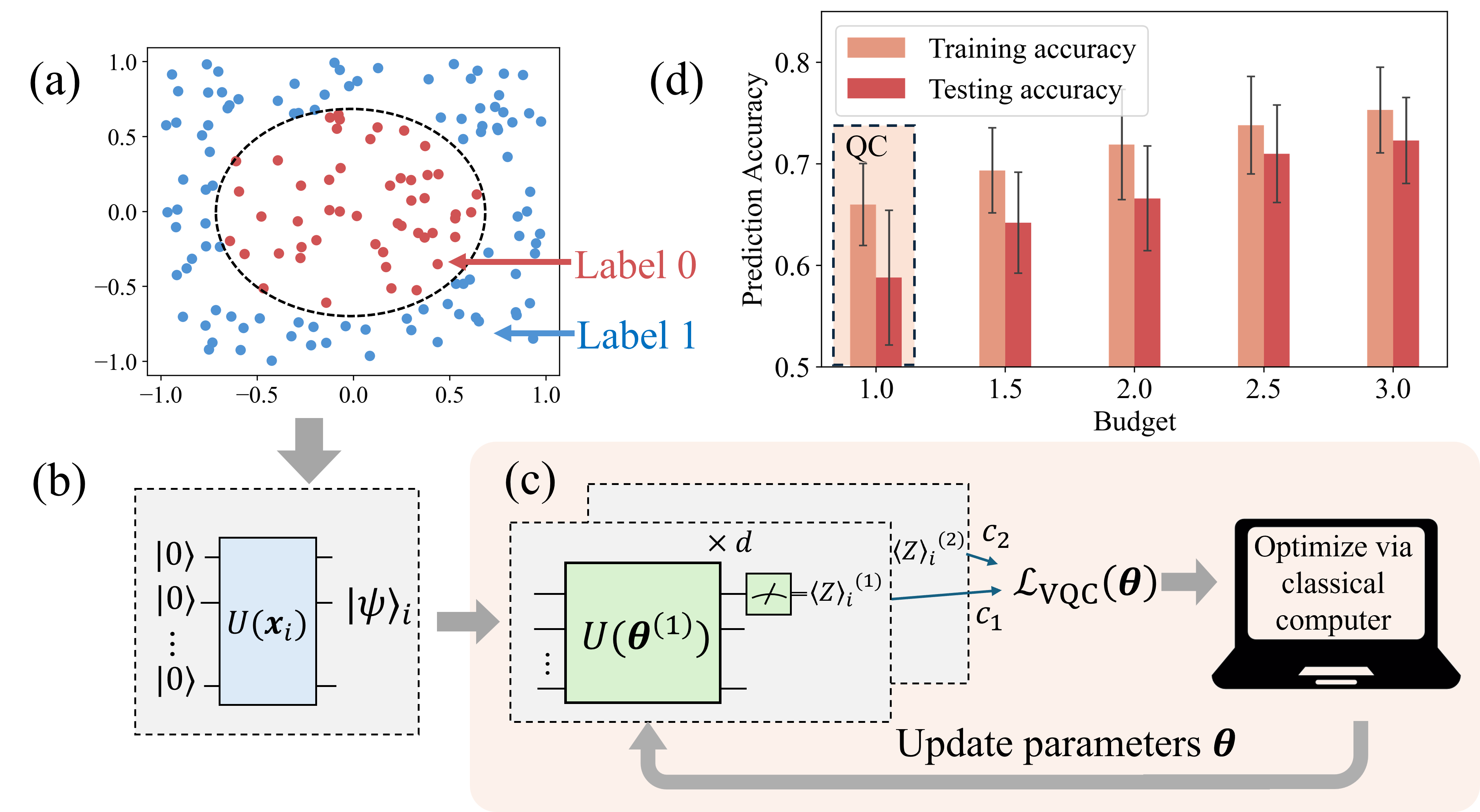}
    \caption{{\bf Virtual quantum classifier.} (a) Training Data. In this example, the input vectors lie in the 2-dimensional plane, and  are labeled with 0 and 1. (b) Data Encoding. The classical data are encoded into quantum state before training the virtual quantum classifier (c) Training virtual quantum classifier. Implement PQCs and make measurements to estimate the loss function on the quantum devices. Then optimize the loss function with classical computers by tuning the parameters in the PQCs. After iterations, the loss function converge to its minimum and out put the optimized virtual quantum classifier (d) Prediction accuracy by optimized VQC with different sampling overhead budgets. The prediction accuracy with the budget equal to 1.0 corresponds to the conventional quantum classifier.}
    \label{fig:diagram_VQC}
\end{figure*}
%%%%%%%%%%%%%%%%%%%%%%%%%%%%%%%%%%%%%%%%%%%%%%%%%%%%%%%%%%%%
\subsection{Quantum Classification}\label{sec:classification}

Quantum classifiers~\cite{li2022recent,grant2018hierarchical} are a central model of supervised  quantum machine learning.  The input for the classification task is a data set consisting of labeled samples $D = \{\boldsymbol{x}_i, y_i\}_{i=1}^M$ where $M$ is the data size, $\boldsymbol{x}_i$ is a vector, representing  a \textit{data point}, and $y_i$ is the corresponding label. In the binary classification task, the labels are   0 and 1, \textit{i.e.}, $y_i\in\{0,1\}$. The first step for the quantum classifier is to encode these input data  into quantum states.  The details on how the encoding is done are summarized  in Appendix~\ref{app:data_encoding}.  Once the encoding has been completed, the classification problem can be viewed as the task of  assigning  an  appropriate label to an unseen quantum state, based  on the information gained from previous training data.

 Here, we apply the VQNN framework to the classification task,  developing  a new algorithm called the \textit{virtual quantum classifier (VQC)}.  Compared to conventional quantum classifiers, the VQC offers an advantage in the presence of unknown quantum noise affecting the encoding operations, as it can suitably adapt its classification policy.

Within the virtual quantum classifier paradigm, the classifier $f_{\boldsymbol{\theta}}$, which is a combination of QNNs with parameters $\boldsymbol{\theta}$ and measurements, is utilized to predict the label $\Tilde{y_i}$. A good quantum classifier $f_{\boldsymbol{\theta}}$ should map data to their corresponding labels as accurately as possible $f_{\boldsymbol{\theta}}(\boldsymbol{x}_i)\rightarrow y_i$. Therefore, in the training procedure, we use the average distance between the actual label $y_i$ and the predicted label $\Tilde{y_i}$ as the loss function, which is
\begin{equation}
    \cL_{\rm VQC}(\boldsymbol{\theta}) = \frac{1}{M}\sum_{i=1}^M|\Tilde{y}_i - y_i|^2.
\end{equation}
The detailed procedure of training a virtual quantum classifier is shown in Appendix~\ref{app:VQC}.

A well trained classifier $f_{\boldsymbol{\theta^*}}$ should also predict unseen data accurately. Thus, we need a testing data set to verify the performance of the trained classifier $f_{\boldsymbol{\theta^*}}$. To demonstrate the power of VQC on noisy quantum device, we conduct a numerical experiment. The task is binary classification, and the visualized training data set is shown in Fig.~\ref{fig:diagram_VQC} (a). The training set contains 200 two-dimensional data points, and the red(blue) points represent the labels of 0(1). The quantum gates in the ansatz of VQC are noisy, which are modeled as the ideal gates followed by noisy channels as shown in Fig.~\ref{fig:Ansatz} (b), and the depth is set as 2. Specifically, the single-qubit rotation gates is followed by single-qubit depolarizing channels with noise level $p=0.02$; the two-qubit rotation gates are followed by two-qubit depolarizing channels with noise level $p=0.05$. After training, we randomly generate a testing data set with 100 data points, which have the same features as the training data set. Then we apply the trained classifier $f_{\boldsymbol{\theta}^*}$ to the testing data to verify the performance of the classifier when encountering the unseen data.
The numerical results are shown in Fig.~\ref{fig:diagram_VQC} (d).  The error bar refers to the standard deviation by repeating 10 times.
Note that the results with a budget equal to $1.0$ (red shaded area denoted as QC) correspond to the conventional quantum classifier.
It can be seen that the prediction accuracy for both the training data set and the testing data set increases with the increase of budgets, demonstrating the advantage of VQNNs.

%%%%%%%%%%%%%%%%%%%%%%%%%%%%%%%%%%%%%%%%%%%%%%%%%%%%%%%%%%%%%%%%%%%%%%%%%%%%%%%%%%%%%%%%%%
\subsection{Estimating ground-state energy}\label{sec:VQE}
Estimating the ground state energy of a Hamiltonian is a fundamental task in quantum physics~\cite{anderson1952approximate}, chemistry~\cite{reiher2017elucidating}, and conductivity and magnetism~\cite{lee2006doping}, as it determines the lowest possible energy and thus the stability and physical properties of a system. In quantum computing, the variational quantum eigensolver (VQE) can be used to estimate the ground state energy of a Hamiltonian $H$ on near-term quantum devices, which is an unsupervised learning task. The loss function is $\cL_{VQE}(\boldsymbol{\theta}) = \tr[H \ketbra{\psi(\boldsymbol{\theta})}{\psi(\boldsymbol{\theta})}]$, where one aims to minimize the expectation value of the Hamiltonian $H$ over a trial state $\ket{\psi(\boldsymbol{\theta})} = U(\boldsymbol{\theta})\ket{0}$ for some parametrized ansatz $U(\boldsymbol{\theta})$ and initial state $\ket{0}$. In practice, the Hamiltonian is typically expressed as a linear combination of tensor products of Pauli operators, $H=\sum_k h_k P_k$, where $h_k$ are real coefficients. Consequently, the loss function $\cL_{VQE}(\boldsymbol{\theta})$ can be evaluated as a weighted sum of the expectation values of the Pauli operators, which can be estimated on quantum devices.

When implementing VQE on quantum devices, however, noise is unavoidable, corrupting the computing and preventing us from accessing the true ground state energy. Here, we exploit the VQNN framework and propose a new algorithm called VVQE to solve this problem.

\subsubsection{Loss function}
We consider the scenario where two branches of PQCs are utilized, and the sampling overhead budget $\gamma_b=|c_1|+|c_2|$ is given, where $c_1=(1+\gamma_b)/2$ and $c_2=(1-\gamma_b)/2$ are real numbers. The output average state is $\sigma = c_1\rho_1+c_2\rho_2$, and $\rho_i=U_i (\boldsymbol{\theta}^{(i)})\ketbra{0}{0} U_i^\dagger(\boldsymbol{\theta}^{(i)}), i\in\{1,2\}$. In this task, we aim to obtain the lowest energy of the Hamiltonian $H$. In the conventional VQE algorithm, it solves an optimization problem of $E^*=\min_\rho\tr[H\rho]$, where $\rho$ is a quantum state. The constraint of the quantum state guarantees that the optimal value $E^*$ is the lowest energy of the Hamiltonian $H$, and the corresponding state $\rho^*$ is the ground state.

In our case, $\sigma$ is the averaged state, which is a general Hermitian operator. If we simply set the loss function as $\tr[H\sigma]$, then the optimal value will be $c_2 E_\infty + c_1 E_0$ if $\gamma_b \ge 1$, where $E_0$ and $E_\infty$ are the minimum and maximum energies of Hamiltonian $H$, respectively. Therefore, we should force the averaged state $\sigma$ as a valid state,
% Therefore, $\tr[H\sigma]$ should be minimized. Here we have to note that the averaged state $\sigma$ is supposed to be a valid quantum state; otherwise, the minimized quantity $\tr[H\sigma]$ is meaningless. Therefore, $\sigma$ should satisfy two constraints: unit trace and positive semi-definite,
i.e., $\tr[\sigma]=1$ and $\sigma \ge 0$. For the unit trace constraint, it is satisfied directly by setting $c_2 = 1 - c_1$. For the positive semi-definite constraint, we could take the negativity of the average state into the loss function $\tr[|\sigma|-\sigma]/2 = \sum_{\lambda_i\le0}|\lambda_i|$, which is the summation of all the absolute value of the negative eigenvalues $\lambda_i$ of the average states $\sigma$. When minimizing the negativity, it will converge to 0, making all eigenvalues of the average state nonnegative, i.e., $\sigma$ is positive semidefinite.  Thus, the natural choice of the loss function is
\begin{equation}
    \cL_{\rm VVQE}(\boldsymbol{\theta}) = \tr[H\sigma] + \alpha \frac{\tr[|\sigma|-\sigma]}{2},
\end{equation}
where $\alpha$ is the penalty strength. However, estimating the negativity of a quantum state on a quantum device is generally hard, implying that such a loss function is not practical. Here, instead of utilizing the negativity directly, we approximate it with a $K$-truncated Chebyshev expansion. Recalling the  expression of the normalized average state $\Tilde{\sigma}=\sigma/|c_1|$ (so that $\sigma=c_1\,\Tilde{\sigma}$, with $c_1=|c_1|>0$), the negativity reads
\begin{equation}
    \frac{\tr[|\sigma|-\sigma]}{2}=c_1\,\tr[f(\Tilde{\sigma})]\,,\qquad f(\Tilde{\sigma}):=\frac{|\Tilde{\sigma}|-\Tilde{\sigma}}{2}\, .
\end{equation}
The details of the Chebyshev approximation of $f(\Tilde{\sigma})$ by the truncated polynomial $p_K(\Tilde{\sigma})$ are given in Appendix~\ref{app:chebyshev}. Approximating $\tr[f(\Tilde{\sigma})]\simeq\tr[p_K(\Tilde{\sigma})]$ then leads to the loss function

\begin{align}\label{eq:loss_VVQE}
    \cL_{\rm VVQE}(\boldsymbol{\theta}) &= \tr[H\sigma] + \alpha\, c_1\,\tr[p_K(\Tilde{\sigma})] \nonumber\\
    &= c_1\tr[H\rho_1] + c_2\tr[H\rho_2] + \alpha\, c_1\,\tr[p_K(\Tilde{\sigma})]\, .
\end{align}

With this loss function, we estimate the ground-state energy with the procedure shown in Fig.~\ref{fig:workflow}. The detailed processes are shown in Appendix~\ref{app:VVQE}.

\subsubsection{Estimating the ground-state energy of $\rm HeH^+$}
Here we conduct the numerical experiment to estimate the ground-state energy of $\rm HeH^+$ on a noisy quantum device with the VVQE algorithm. Conduct Algorithm~\ref{algo:VVQE} by setting the noisy PQCs as shown in Fig.~\ref{fig:Ansatz} (b) with depth $d=5$. The truncation number $K$ is set as 100. The PQCs noise is set as single-qubit depolarizing and two-qubit depolarizing noise with noise level $p=0.01$ and $p=0.02$, respectively.

\begin{figure}[h]
    \centering
    \includegraphics[width=\linewidth]{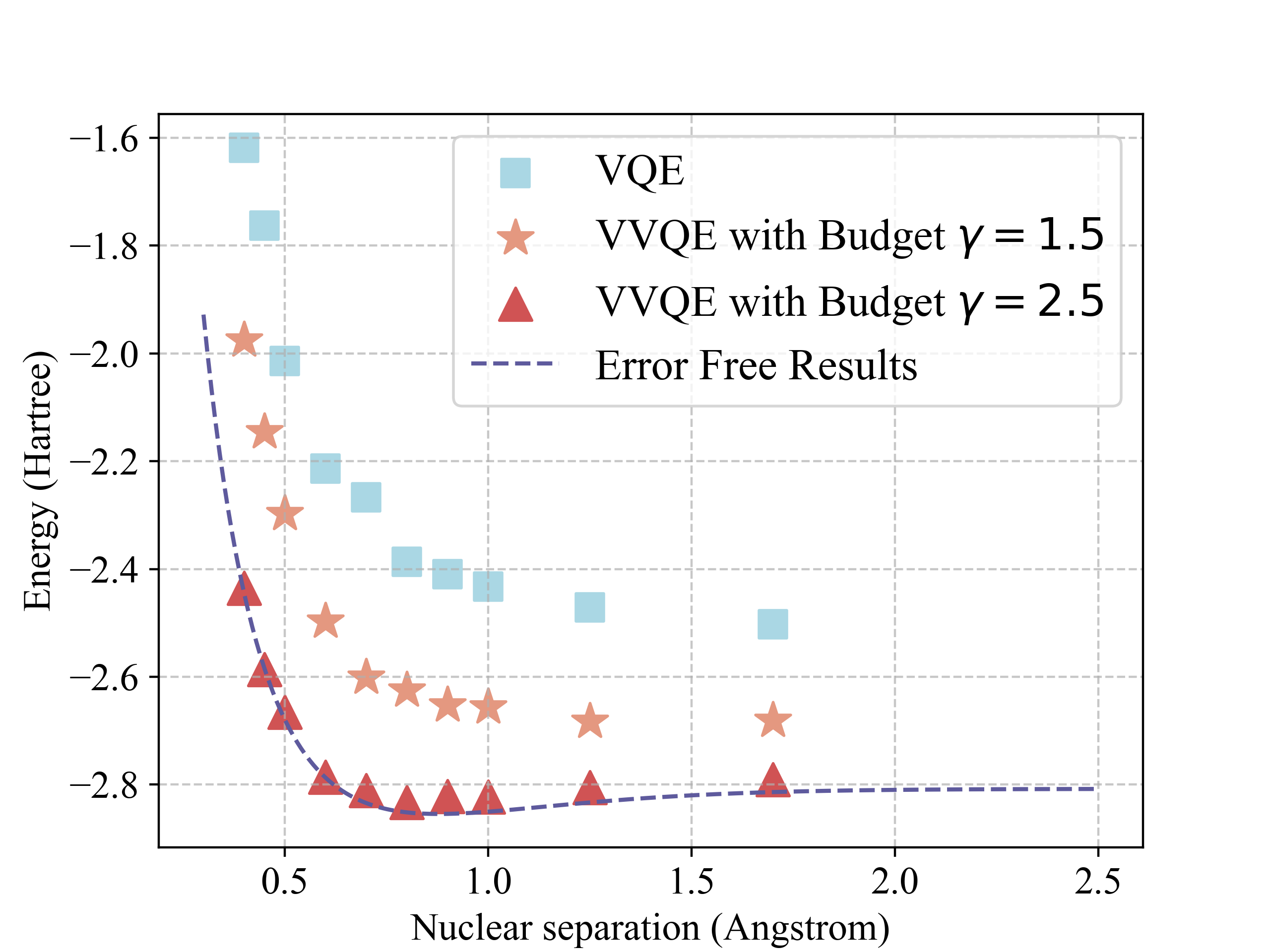}
    \caption{{\bf Ground state energy of the molecule $\rm HeH^+$ estimated via the VVQE algorithm.} The dashed line shows the exact values of the ground state energy as a function of the nuclear separation.  The orange stars and red triangles show the energy values found by our VVQE algorithm  for different values of the sampling budgets, set to $\gamma=1.5$ and $\gamma =2.5$,  respectively. The light blue squares show the energies computed with the conventional VQE, which is equivalent to a VVQE with unit budget $\gamma=1$.}
    \label{fig:VVQE_HeH+}
\end{figure}

The numerical results are shown in Fig.~\ref{fig:VVQE_HeH+}. The dark blue dashed line is the error-free ground state energy. The light blue squares correspond to the energy estimation via conventional VQE with noisy circuits. The orange stars and red triangles are obtained by the proposed virtual VQE with budgets $\gamma=1.5$ and $\gamma=2.5$, respectively. Obviously, an accurate ground state energy can be estimated with a reasonably large budget.

%%%%%%%%%%%%%%%%%%%%%%%%%%%%%%%%%%%%%%%%%%%%%%%%%%%%%%%%%%%%
\section{Discussion and conclusions}\label{sec:discussion}
In this work we introduced the framework of virtual quantum neural networks, leveraging random sampling and classical data processing to enhance the expressivity  and performance of quantum circuits in the NISQ regime. To illustrate the benefits of our framework, we  applied it  to three representative tasks, including both supervised and unsupervised learning. Compared with the conventional QNN, VQNN shows advantages in two aspects: enhanced expressivity and improved noise robustness.

The enhanced expressivity arises from an enlarged optimization space. Conventional QNNs restrict the optimization to a single parametric family of quantum circuits, represented by completely positive maps.  VQNNs, instead, use sampling over multiple quantum circuits to achieve a larger optimization space, mathematically described by a family of Hermitian-preserving maps.  Overall, one can view VQNNs as a way to overcome some of the limitations of the existing quantum hardware by investing additional resources in sampling  and classical computation at the training stage.

A promising direction is the application of the VQNN framework to scenarios that require long-range quantum entanglement. In a conventional QNN, establishing long-range entanglement requires  quantum circuits of large depth~\cite{wen2013topological,chen2010local}, which however tend to be affected by barren plateaus. Shallow circuits, instead, are gradient-healthy, but their expressive power is limited by their causal cones,  which cannot support the generation of long-range entanglement. In contrast, linear combinations of shallow quantum circuits could be used to virtually simulate  long-range entanglement~\cite{mitarai2021constructing, mitarai2021overhead, piveteau2023circuit}. This intuition  is made rigorous by the observation that arbitrary non-local quantum operations can be simulated by  linear combinations of local operations and classical communication, as shown in  Appendix~\ref{sec:simulate non-local gate}.

% The advantage of expressivity of VQNN can be partially explained by two aspects: the searching space and the long-range entanglement. \xq{which part of the results implies that long-range entanglement plays an important role?} \bc{Maybe we can cite Ma's paper, the LOCC one?}\sout{The explanation for the first point is trivial.} VQNN relaxes the searching space from CPTP to HPTP, and it is natural to expect VQNN to learn the more complicated relations. For the second point, the long-range entanglement structures play an essential role in many quantum information processing scenarios.   Therefore, the VQNN could simulate more entangled circuits than the conventional QNN with shallow circuits. \xq{this sounds similar to the enlargement of the search space}\bc{a little bit?}

Another key feature is the robustness of VQNNs against hardware noise. 
In conventional QNN, noise accumulates with circuit depth, leading to a loss of signal-to-noise ratio. In contrast, VQNN can  partially cancel noise  by creating linear combinations of multiple quantum circuits. Since the optimal coefficients in the linear combination  are learned in the presence of noise, the optimization process naturally functions as a variational protocol for  probabilistic error cancellation~\cite{temme2017error,jiang2021physical,zhao2023information}.
%as discussed in Sec.~\ref{sec:error mitigation}.

The above observations  suggest that the sampling and classical postprocessing serve VQNNs with a dual purpose: on the one hand, they act as a resource to bypass the connectivity constraints of the available quantum hardware and, on the other hand, they act as a buffer that absorbs  noise.  In this way,  the model’s expressivity is freed from some of the key limitations of the existing hardware, including limitations  both at the logical level (restricted set of logical operations) and at the physical level (decoherence and hardware imperfections). 

%In future work, it would be interesting to apply the VQNN to other QML tasks. Besides, it is important to study the BP problem and expressivity theoretically. If having access to the quantum hardware, it would also be great to implement the VQNN algorithms and to show the quantum advantage.

%%%%%%%%%%%%%%%%%%%%%%%%%%%%%%%%%%%%%%%%%%%%%%%%%%%%%%%%%%%%

\textbf{Acknowledgements.--}B. Z would like to thank Mingrui Jing, Xiao Shi and Ya-dong Wu for valuable discussions.   This work was supported by the Chinese Ministry of Science and Technology (MOST) through grant 2023ZD0300600, and by the Hong Kong Research Grant Council (RGC) through grants  SRFS2021-7S02  and R7035-21F, and the State Key Laboratory of Quantum Information Technologies and Materials.  Yingzhou Li is supported in part by the Scientific Research Innovation Capability Support Project for Young Faculty under grant SRICSPYF-ZY2025159, National Natural Science Foundation of China under grant 12526211, the Science and Technology Commission of Shanghai Municipality under grants 22TQ017. Yinan Li was supported by the National Key Research and Development Program of China under Grant 2024YFE0102500, the Fundamental Research Funds for the Central Universities under Grant 2042025kf0023, the National Nature Science Foundation of China under Grant 62302346 and the Hubei Provincial Nature Science Foundation of China under Grant 2024AFA045.

\textbf{Data availability.--}The data that support the findings of this article are openly available~\cite{codes}.
%%%%%%%%%%%%%%%%%%%%%%%%%%%%%%%%%%%%%%%%%%%%%%%%%%%%%%%%%%%%
% Bibliography
%%%%%%%%%%%%%%%%%%%%%%%%%%%%%%%%%%%%%%%%%%%%%%%%%%%%%%%%%%%%
% \bibliographystyle{apsrev4-1}
\bibliography{ref}

%%%%%%%%%%%%%%%%%%%%%%%%%%%%%%%%%%%%%%%%%%%%%%%%%%%%%%%%%%%%
%%%%%%%%%%%%%%%%%%%%%%%%%%%%%%%%%%%%%%%%%%%%%%%%%%%%%%%%%%%%
% \clearpage

\vspace{2cm}
\onecolumngrid
\vspace{2cm}
\begin{center}
{\textbf{\large Appendix for Virtual Quantum Neural Networks}}
\end{center}

\appendix

%%%%%%%%%%%%%%%%%%%%%%%%%%%%%%%%%%%%%%%%%%%%%%%%%%%%%%%%%%%%
% \renewcommand{\appendixname}{Supplementary Note}

%%%%%%%%%%%%%%%%%%%%%%%%%%%%%%%%%%%%%%%%%%%%%%%%%%%%%%%%%%%%%%%%%%%%%%%%%%%

%%%%%%%%%%%%%%%%%%%%%%%%%%%%%%%%%%%%%%%%%%%%%%%%%%%%%%%%%%%%%%%%%%%%%%%%%%%%%
\section{Analysis of Barren Plateaus}\label{app:BP}
The problem of barren plateaus (BPs)~\cite{larocca2025barren} arises when the gradients of the loss function $\cL(\boldsymbol{\theta})$ vanish exponentially as the number of qubits $n$ grows. Mathematically, a BP corresponds to the condition   $\text{Var}[\partial_{\theta_\mu} \cL(\boldsymbol{\theta})]\in \cO({1}/{b^n})$,
for some constant $b>1$, and for some (not necessarily all) parameters
$\theta_\mu\in \boldsymbol{\theta}$.
For many common parameterized quantum circuit architectures, the parameter space becomes uniformly explored across the exponentially large Hilbert space, making the cost landscape overwhelmingly flat and featureless. Consequently, optimizing the parameters becomes intractable as it requires an exponentially large number of measurements to discern a reliable optimization direction, posing a fundamental challenge to training scalable QNNs.

We now analyze how the VQNN framework affects this phenomenon. For simplicity, we take the function $f$ entering the loss function Eq.~\eqref{eq:general_cost_func} to be the identity map, so that the VQNN loss becomes
\begin{equation}
    \cL_{\rm VQNN}(\boldsymbol{\theta,c}) = \sum_{i=1}^N c_i \, \tr\left[O\, U_i(\boldsymbol{\theta}^{(i)})\rho_0 U_i^\dagger(\boldsymbol{\theta}^{(i)})\right] =: \sum_{i=1}^N c_i \, \cL_i(\boldsymbol{\theta}^{(i)}) \, ,
\end{equation}
where $\cL_i(\boldsymbol{\theta}^{(i)})$ is the loss function of the $i$-th branch. The gradient with respect to a parameter $\theta_\mu^{(i)}$ of the $i$-th branch is therefore $\partial_{\theta_\mu^{(i)}} \cL_{\rm VQNN}(\boldsymbol{\theta,c}) = c_i \, \partial_{\theta_\mu} \cL_i(\boldsymbol{\theta}^{(i)})$, and, since this gradient depends only on the $i$-th branch, its variance factorizes as
\begin{equation}
    \text{Var}\left[\partial_{\theta_\mu^{(i)}} \cL_{\rm VQNN}(\boldsymbol{\theta,c})\right] = c_i^2 \, \text{Var}\left[\partial_{\theta_\mu} \cL_i(\boldsymbol{\theta}^{(i)})\right] \, .
\end{equation}
Suppose now that each branch, taken on its own, exhibits a barren plateau, i.e., $\text{Var}[\partial_{\theta_\mu} \cL_i] \in \cO(1/b^n)$. The above relation then gives $\text{Var}[\partial_{\theta_\mu^{(i)}} \cL_{\rm VQNN}] = c_i^2 \, \cO(1/b^n)$. As long as the coefficients are bounded, $|c_i| = \cO(1)$, which is guaranteed by a bounded sampling overhead $\gamma = \sum_i |c_i|$, the variance remains exponentially small in $n$; the coefficients only rescale it by a constant factor. In other words, a VQNN with bounded sampling overhead does not escape the barren plateau affecting its constituent branches, but it also does not aggravate it: the variance of each branch's gradient is of the same exponential order as for a standard QNN. Thus, the additional expressivity of the VQNN framework does not come at the price of a worse barren-plateau problem. Conversely, substantially mitigating a barren plateau would require $|c_i|$ to grow exponentially with $n$, which in turn entails an exponentially large sampling overhead and is therefore outside the regime of practical interest.

\section{Derivation of Eq.~\eqref{bbb}}\label{app:derivatino_of_eq}
In this Appendix we provide the derivation of Eq.~\eqref{bbb} in the main text.

The starting point is to reformulate the optimization problem in Eq.~\eqref{bbb} of the main text using the Choi correspondence between quantum processes and bipartite quantum states~\cite{choi1975completely}.  Mathematically, the Choi operator  of a  linear map $\cN:  L({\cal H}_A)  \to  L({\cal H}_B)$ is the operator $J^\cN \in L({\cal H}_A \otimes {\cal H}_B)$ defined as
\begin{equation}\label{choioperator}
    J^\cN = \sum_{i,j=0}^{d_A-1}\ketbra{i}{j} \ox \cN(\ketbra{i}{j}) \, ,
\end{equation}
where $d_A$ is the dimension of the input system, and $\{  |i\rangle\}_{i=0}^{d_A-1}$ is a fixed orthonormal basis for ${\cal H}_A \simeq \mathbb{C}^{d_A}$.
%For a CPTP map $\cN$, the Choi matrix is proportional to a  quantum state; explicitly, the Choi \textit{state} of  $\cN$ is the density matrix
 %   $\Tilde{J}^{\cN}_{AB} := \frac{1}{d_A} J^{\cN}_{AB}$.

Eq. (\ref{choioperator}) can be rewritten in a more compact way by introducing the following notations. For a quantum system $A'$ with the same dimension as $A$, we denote by ${\cal H}_A \otimes {\cal H}_{A'}$ by $|\Phi^+ \>_{AA'}  :  = \sum_i |i\>_A  \otimes |i\>_{A'}/\sqrt{d_A} $ the canonical maximally entangled state of $AA'$, and we will use the notation $\Phi^+_{AA'}  : = |\Phi^+\rangle\langle \Phi^+|_{AA'}$ for the corresponding projector.  We will also use subscripts to indicate the systems on which map act, \textit{e.g.} writing $\cN_{A\rightarrow B}$ to indicate that $A$ and $B$ are the input and output systems of the map $\cN$, respectively.    With this notation, the Choi matrix can be written as $  J^\cN_{AB} =   d_A \,  \cN_{A'\to B} (\Phi^+_{AA'})$, where we omitted the identity map $\cI_{A\rightarrow A}$.

 We are now ready to reformulate Eq.~(\ref{eq:original_problem})  in terms of the Choi operator.  First, note that the error mitigation condition    $\cN^{-1}=\sum_ic_i\cC_i$ is equivalent to  $\cI  =   \cN^{-1} \circ \cN =  \sum_ic_i~ \cC_i\circ\cN$, which can be written as $\sum_i  {J}_{\cC_i\circ\cN} = {J}_\cI  $. By Eq. (\ref{choioperator}), this condition is equivalent to
\begin{align}\label{aaa}
    \sum_i c_i ~\cC_{i, B\rightarrow A''} \circ \cN_{A'\rightarrow B}(\Phi^+_{AA'}) = \Phi^+_{AA''},
\end{align}
where $A''$ is another quantum system with $d_A$-dimensional Hilbert space.

In general, each quantum channel $\cC_i: L({\cal H}_B) \to L({\cal H}_A)$ can be realized by performing a unitary channel $\cU_i:  L({\cal H}_B \otimes {\cal H}_R) \to L({\cal H}_A \otimes {\cal H}_{R'})$ on a larger system including an auxiliary input system  $R$  and an auxiliary output system $R'$, of dimensions $d_R$ and $d_{R'}$ satisfying the conditions $d_B d_R  =  d_Ad_{R'}$ and $d_{R'} \le d_Ad_B$.   In this unitary realization, the action of the channel $\cC_i$ on a generic input state $\rho$ is decomposed as
\begin{align}\label{unitaryrealization}\cC_i (\rho)  = \tr_{R'} [  \cU_i  (\rho  \otimes |0\>\<0|)  ] \,,
\end{align}where $|0\> \in {\cal H}_R$ is a fixed initial state of system $R$.

Adopting the unitary realization (\ref{unitaryrealization}), the error mitigation condition (\ref{aaa}) becomes
\begin{align}\label{eq:stinespring}
    \sum_i c_i \tr_{R'}[\cU_{i, BR\rightarrow A''R'}(\cN_{A'\rightarrow B}(\Phi^+_{AA'})\ox\ketbra{0}{0}_{R}])] = \Phi^+_{AA''}.
\end{align}
We now introduce  the channel $\cD_i:  L({\cal H}_{A'} \otimes {\cal H_{R}})\to L({\cal H}_{A''})$, defined as
\begin{align}\cD_{i,  A'R\to S''} :  = \tr_{R'} \circ  \, \cU_{i,  BR  \to  A'' R'}  \circ  \cN_{A'\to B} \,.
\end{align}
With this definition,  Eq. (\ref{eq:stinespring}) becomes
\begin{align}
\sum_i c_i\, \cD_{i,  A'R \to A''}    (  \Phi^+_{AA'} \otimes |0\>\<0|_R)  =  \Phi^+_{AA''} \,,
\end{align}
and the optimization problem in Eq. (\ref{eq:original_problem}) of the main text becomes
\begin{align}
    \min_{c_i, U_i}  \Big\{\sum_i |c_i|\, \Big|\, \sum_i c_i \cD_{i,  A'R\to A''}    (\Phi^+_{A'A} \otimes |0\>\<0|_R)  = \Phi^+_{A''A} \,  , c_i  \in \mathbb{R}  \, , ~ \cU_i \in \text{Unitary}\Big\} \, ,
\end{align}
where  $\text{Unitary}$ denotes the set of all unitary CPTP maps with input $L(  {\cal H}_B \otimes  {\cal H}_{R'})$ and output  $L(  {\cal H}_{A''} \otimes  {\cal H}_{R'})$.   This concludes the proof of Eq.  (\ref{bbb}) in the main text.

%%%%%%%%%%%%%%%%%%%%%%%%%%%%%%%%%%%%%%%%%%%%%%%%%%%%%%%%%%%%%%%%%%%%%%%%%%%%%%%%
\section{Algorithms for variational quantum error mitigation} \label{app:VQEM}

\begin{algorithm}[H]
\renewcommand{\algorithmicrequire}{\textbf{Input:}}
    \renewcommand{\algorithmicensure}{\textbf{Output:}}
    \begin{algorithmic}[1]
        \REQUIRE Noise channel $\cN$, real coefficients $\boldsymbol c=\{c_i\}_{i=1}^N$, penalty strength $\alpha$,  $N$-parameterized quantum channels $\{\cC_i(\boldsymbol{\theta}_i)\}_{i=1}^N$, number of iterations ITR.
        \ENSURE optimized sampling overhead $\gamma'$, quantum protocol to mitigate the noise $\cN$.
        \STATE Initialize the parameters  $\boldsymbol{\theta}$ and the real coefficient $c$.
        \FOR {itr = 1, $\dots$, ITR}
            \STATE Prepare $N$-copies of maximally entangled state $\Phi^+_{AB}$ and apply the quantum noise $\cN$ on system $B$.
            \STATE Apply quantum channels $\cC_i(\boldsymbol{\theta}_i)$ in parallel on the quantum systems $BR$, and obtain states $\{\sigma_i\}_{i=1}^N$.
            \STATE Compute the sampling overhead $\gamma'=\|\boldsymbol{c}\|_1$.
            \STATE Measure $\tr[\sigma_i\sigma_j], \, \forall\, i,j = 1,\cdots,N$.
            \STATE Measure $\tr[\Phi^+\sigma_i], \, \forall\, i= 1,\cdots,N$.
            \STATE Compute the loss function $\cL^{(1)}_{\rm VQEM}(\boldsymbol{\theta}, \boldsymbol{c})$ as shown in Eq.~\eqref{eq:VQEM_loss_function_1}.
            \STATE Update the parameters $\boldsymbol{\theta}$ and coefficients $\boldsymbol{c}$ to minimize the loss function.
        \ENDFOR
        \STATE Output the optimized sampling overhead $\gamma'$ and the corresponding quantum mitigation protocol.
    \end{algorithmic}

\caption{Variational quantum error mitigation (VQEM)}
\label{algo:VQEM}
\end{algorithm}

\begin{algorithm}[H]
    \renewcommand{\algorithmicrequire}{\textbf{Input:}}
    \renewcommand{\algorithmicensure}{\textbf{Output:}}
    \begin{algorithmic}[1]
        \REQUIRE Noise channel $\cN$, fixed coefficients $\boldsymbol{c}=\{c_i\}_{i=1}^N$ with sampling overhead budget $\gamma_b$, penalty strength $\alpha$, $N$-parameterized quantum channels $\{\cC(\boldsymbol{\theta}_i)\}_{i=1}^N$, number of iterations ITR.
        \ENSURE Quantum protocol to mitigate the noise
        $\cN$.
        \STATE Initialize the parameters $\boldsymbol{\theta}$.
        \FOR {itr = 1, $\dots$, ITR}
            \STATE Prepare $N$-copies of maximally entangled state $\Phi^+_{AB}$ and apply the quantum noise $\cN$ on system $B$.
            \STATE Apply quantum channels $\cC_i(\boldsymbol{\theta}_i)$ in parallel on the quantum systems $BR$, and obtain states $\{\sigma_i\}_{i=1}^N$.
            \STATE Measure $\tr[\sigma_i\sigma_j], \, \forall\, i,j = 1,\cdots,N$.
            \STATE Measure $\tr[\Phi^+\sigma_i], \, \forall\, i= 1,\cdots,N$.
            \STATE Compute the loss function $\cL^{(2)}_{\rm VQEM}(\boldsymbol{\theta})$ as shown in Eq.~\eqref{eq:VQEM_loss_function_2}.
            \STATE Update the parameters $\boldsymbol{\theta}$ and coefficients $\boldsymbol{c}$ to minimize the loss function.
        \ENDFOR
        \STATE Output the optimized quantum mitigation protocol.
    \end{algorithmic}
\caption{Variational quantum error mitigation (VQEM) with budget}
\label{algo:VQEM_2}
\end{algorithm}

%%%%%%%%%%%%%%%%%%%%%%%%%%%%%%%%%%%%%%%%%%%%%%%%%%%%%%%%%%%%%%%%%%%%%%%%%%%%%%%%%%%%
\section{SDP for channel inverse $\cN^{-1}$}\label{app:SDP}
The inverse channel $\cN^{-1}$ is generally not a CPTP map but an HPTP map, which cannot be implemented on a quantum device directly. Therefore, we have to decompose the inverse channel into a linear combination of CPTP maps,
\begin{equation}
    \cN^{-1} = \sum_i c_i \cC_i,
\end{equation}
where $c_i$ are the real coefficients and $\sum_i c_i=1$ and $\cC_i$ are the CPTP maps. With the decomposition, one can simulate the unphysical map.

According to Ref.~\cite{jiang2021physical}, any HPTP map can be represented with only two CPTP maps
\begin{lemma}\label{lemma:QPD_2_term}~\cite{jiang2021physical, zhao2023information}
    Let $\cM$ be an HPTP map, then there exist two real numbers $c_1\ge0, c_2\ge0$ and two CPTP maps $\cC_1, \cC_2$ such that
    \begin{equation}
        \cM = c_1 \cC_1 - c_2\cC_2.
    \end{equation}
\end{lemma}
The decomposition can be simplified into $\cN^{-1} = c1 \cC_1 - c_2\cC_2$.

Note that the decomposition of an unphysical map is not unique, and the optimal one is defined by the one with minimal sampling overhead. Then, the optimal decomposition problem can be written as a semidefinite problem as shown in the following. For a given noisy channel $\cN$, the error mitigation protocol with minimum sampling overhead can be calculated by the following SDP
\begin{subequations}
    \begin{align}
        \min\quad& c_1 + c_2\\
        \text{subject to}\quad & J_{\cN^{-1}} = J_{\cC_1} - J_{\cC_2},\\
        & J_{\cC_1} \ge 0, \, \tr_{out}[J_{\cC_1}] = c_1 I_{in},\\
        & J_{\cC_2} \ge 0, \, \tr_{out}[J_{\cC_2}] = c_2 I_{in},
    \end{align}
\end{subequations}
where $J_{\cN^{-1}}$ refers to the Choi matrix of the noise inverse channel $\cN^{-1}$.

\section{Data encoding}\label{app:data_encoding}
In quantum classifier~\cite{havlivcek2019supervised}, the input data is classical, which cannot be utilized by quantum devices. Therefore, we have to transfer the classical data into quantum data before classifying the data. Such a process is called data encoding. There are various ways to encode classical data, such as angle encoding~\cite{breuer2002theory}, basis encoding~\cite{breuer2002theory, larose2020robust}, amplitude encoding~\cite{breuer2002theory}, and Hamiltonian evolution ansatz encoding~\cite{huang2021power}. In this work we adopt angle encoding for its simplicity. Specifically, for the classical feature vector $\boldsymbol{x}=[x_1,\cdots,x_i,\cdots , x_M]^T$, the encoded state is \begin{equation}
    \ket{\psi}_{\boldsymbol{x}}=\bigotimes_{i=1}^M (\cos(x_i)\ket{0} + \sin(x_i)\ket{1}).
\end{equation}
Such an encoded state can be realized by a fixed depth circuit with local rotation gates.
%%%%%%%%%%%%%%%%%%%%%%%%%%%%%%%%%%%%%%%%%%%%%%%
\section{Algorithm for virtual quantum classifier}\label{app:VQC}
 \begin{algorithm}[h]
    \renewcommand{\algorithmicrequire}{\textbf{Input:}}
    \renewcommand{\algorithmicensure}{\textbf{Output:}}
    \begin{algorithmic}[1]
        \REQUIRE Training data set $D =\{\boldsymbol{x}_i, y_i\}_{i=1}^M$, sampling overhead budget $\gamma_b$, two parameterized quantum circuits (PQCs) $U(\boldsymbol{\theta}^{(1)}), U(\boldsymbol{\theta}^{(2)})$, number of iterations ITR.
        \ENSURE Optimized virtual quantum classifier $f_{\boldsymbol{\theta^*}}$.
        \STATE Encode the classical data $\boldsymbol{x}_i$ into quantum states $\ket{\psi_i}$ for all $i$.
        \STATE Initialize the parameters $\boldsymbol{\theta}^{(1)}$ and $\boldsymbol{\theta}^{(2)}$.
        \FOR {itr = 1, $\dots$, ITR}
            \FOR {i = 1, $\dots$, M}
                \STATE Apply $U(\boldsymbol{\theta}^{(1)})$ on the encoded quantum states $\ket{\psi_i}$.
                \STATE Make measurement on the first qubit and obtain $\langle Z \rangle_i^{(1)}$.
                \STATE Similarly, apply $U(\boldsymbol{\theta}^{(2)})$ on the state $\ket{\psi_i}$, measure the first qubit and obtain $\langle Z \rangle_i^{(2)}$.
                \STATE Calculate the prediction $\Tilde{y}_i=\frac{1}{2}(1+c_1\langle Z \rangle_i^{(1)} + c_2 \langle Z \rangle_i^{(2)})$, where $c_1=\frac{1+\gamma_b}{2}$ and $c_2=\frac{1-\gamma_b}{2}$.
            \ENDFOR
            \STATE Compute the loss function $\cL_{\rm VQC}(\boldsymbol{\theta}) = \frac{1}{M}\sum_{i=1}^M|\Tilde{y}_i - y_i|^2$.
            \STATE Minimize the loss function and update the parameters $\boldsymbol{\theta}$.
            \STATE The cost converges to its minimum and the corresponding parameters are $\boldsymbol{\theta^*}$.
        \ENDFOR
        \STATE Output the optimized virtual quantum classifier $f_{\boldsymbol{\theta^*}}$.
    \end{algorithmic}
\caption{Virtual quantum classifier (VQC)}
\label{algo:VQC}
\end{algorithm}

%%%%%%%%%%%%%%%%%%%%%%%%%%%%%%%%%%%%%%%%%%%%%%%%%%%%%%%%%%%%%%%%%%%%%%%%%%%%%%%%%%
\section{Chebyshev approximation}\label{app:chebyshev}

For a given budget $\gamma_b\ge1$, the coefficients are $c_1 = \frac{1+\gamma_b}{2}$ and $c_2=\frac{1-\gamma_b}{2}$. The spectrum of the averaged state $\sigma=c_1\rho_1+c_2\rho_2$ is contained in $[-|c_2|, |c_1|]$, and the normalized averaged state is denoted as $\Tilde{\sigma} = \frac{\sigma}{|c_1|}$. Our target is to approximate the function
\begin{equation}
    f(\Tilde \sigma) = \frac{|\Tilde \sigma| - \Tilde \sigma}{2}
\end{equation}
with Chebyshev approximation. The operator absolute value admits the Chebyshev expansion
\begin{equation}
    |\Tilde \sigma| = \frac{2}{\pi} \mathbb{I} + \sum_{k=1}^\infty\frac{4(-1)^{k+1}}{\pi (4k^2-1)} T_{2k}(\Tilde{\sigma}),
\end{equation}
where $T_n(\sigma)$ denotes the Chebyshev polynomial of degree $n$ evaluated at $\sigma$ via functional calculus.
\begin{equation}
    T_0(\Tilde \sigma)=1,\quad T_1(\Tilde \sigma)=\Tilde \sigma,\quad T_{n+1}(\Tilde \sigma) = 2\Tilde \sigma T_n(\Tilde \sigma) - T_{n-1}(\Tilde \sigma).
\end{equation}
Then we obtain
\begin{equation}
    f(\Tilde{\sigma}) = \frac{|\tilde \sigma| - \Tilde \sigma}{2} = \frac{1}{\pi}\mathbb{I} - \frac{1}{2}\Tilde{\sigma} + \sum_{k=1}^\infty\frac{2(-1)^{k+1}}{\pi (4k^2-1)} T_{2k}(\Tilde{\sigma})
\end{equation}
Define the truncated expansion
\begin{equation}
    p_K(\Tilde \sigma) = \sum_{n=0}^K a_n T_n(\Tilde\sigma),
\end{equation}
with coefficients $a_n$ inherited from the above series. By the spectral theorem,
\begin{equation}
    f(\Tilde \sigma) - p_K(\Tilde \sigma) = \sum_{i}(f(\lambda_i)-p_K(\lambda_i))\ketbra{i}{i}
\end{equation}
where $\{\lambda_i\}$  are the eigenvalues of $\Tilde \sigma$. Using the uniform bound $\|T_n(\Tilde \sigma)\|_\infty\le 1$ valid whenever $\text{spec}(\Tilde \sigma) \subset[-1,1]$, the Chebyshev tail satisfies
\begin{equation}\label{eq:app-one bound}
    \|f(\Tilde \sigma)-p_K(\Tilde \sigma)\|_\infty \le\sum_{n>K} |a_n| = \frac{2}{\pi(K+1)}.
\end{equation}

Let $r=\text{rank}(\Tilde \sigma)$, Then
\begin{align}
    |\tr[f(\Tilde \sigma)] - \tr[p_K(\Tilde \sigma)]| &= |\tr[f(\Tilde \sigma) - p_K(\Tilde \sigma)]|\\
    & \le \sum_{i=1}^r |f(\lambda_i) - p_K(\lambda_i)|\\
    &\le r \|f(\Tilde \sigma)-p_K(\Tilde \sigma)\|_\infty.
\end{align}
Combine it with Eq.~\eqref{eq:app-one bound}, we have
\begin{equation}\label{eq:schebyshev_bound}
    |\tr[f(\Tilde \sigma)] - \tr[p_K(\Tilde \sigma)]| \le \frac{2r}{\pi(K+1)}.
\end{equation}

Note that each Chebyshev Polynomial $T_n(\sigma)$ is a degree-$n$ polynomial in $\Tilde{\sigma}$,
\begin{equation}
    T_n(\Tilde{\sigma}) = \sum_{k=0}^{\lfloor n/2 \rfloor} c_{n,k}\Tilde \sigma ^{n-2k}.
\end{equation}
Therefore,
\begin{equation}
    \tr[T_n(\Tilde{\sigma})] = \sum_{k=0}^{\lfloor n/2 \rfloor} c_{n,k}\tr[\Tilde \sigma ^{n-2k}],
\end{equation}
the Chebyshev polynomial is a linear combination of high-order moments.
According to Ref.~\cite{chen2025simultaneous, shi2025near}, such a quantity can be efficiently estimated with sampling complexity $\cO(k\log(k) /\varepsilon^2)$, where $k$ is the truncation number, and $\varepsilon$ refers to the additive number.
Even on noisy quantum devices, the high-order moments can be retrieved with error mitigation techniques~\cite{zhao2024retrieving,temme2017error}.

From Eq.~\eqref{eq:schebyshev_bound}, the error is upper bounded by $\cO(\frac{r}{K})$. For the average state $\sigma$ with large rank $r$, the truncation number $K$ should increase correspondingly to guarantee the precision of approximation. However, in real case, the increase of the truncation number $K$ is mitigated for three reasons: $^{(1)}$ The upper bound is calculated by the largest error $\|f(\Tilde \sigma) - p_K(\Tilde \sigma)\|_\infty$ times the rank $r$, which is very loose. In general case, such a bound cannot be saturated; $^{(2)}$ In optimization, the absolute value of the objective function need not be precise; it suffices for the gradient direction to be correct. $^{(3)}$ Errors arising from high rank are distributed across numerous eigenvalues. As these errors arise from polynomial oscillations, they tend to partially cancel during tracing (summation) and do not deviate entirely in one direction.

%%%%%%%%%%%%%%%%%%%%%%%%%%%%%%%%%%%%%%%%%%%%%%%%%%%%%%%%%%%%%%%%%%%%%%%%%%%%
\section{Algorithm for virtual variational quantum eigensolver}\label{app:VVQE}

\begin{algorithm}[h!]
    \renewcommand{\algorithmicrequire}{\textbf{Input:}}
    \renewcommand{\algorithmicensure}{\textbf{Output:}}
    \begin{algorithmic}[1]
        \REQUIRE Truncation number $K$, real coefficients $c_1, c_2$, penalty strength $\alpha$, two parameterized quantum circuits (PQCs) $U(\boldsymbol{\theta}^{(1)}), U(\boldsymbol{\theta}^{(2)})$, number of iterations ITR, Hamiltonian $H$.
        \ENSURE The ground-state energy of Hamiltonian $H$.
        \STATE Initialize the quantum system to zero states, and randomly prepare the parameters $\boldsymbol{\theta}^{(1,2)}$.
        \FOR {itr = 1, $\dots$, ITR}
            \STATE Apply $U(\boldsymbol{\theta}^{(1)})$ to zero state and achieve state $\rho_1$.
            \STATE Estimate the expectation value $\tr[H\rho_1]$.
            \STATE Similarly, apply $U(\boldsymbol{\theta}^{(2)})$ to achieve state $\rho_2$, and estimate the expectation value $\tr[H\rho_2]$.
            \STATE Estimate the truncated Chebyshev approximation $p_K$.
            \STATE Compute the loss function $\cL_{\rm VVQE}(\boldsymbol{\theta})$ as shown in Eq.~\eqref{eq:loss_VVQE}.
            \STATE Minimize the loss function and update parameter $\boldsymbol{\theta}$.
        \ENDFOR
        \STATE When the loss function is minimized, we obtain the optimized parameterized quantum circuits $U(\boldsymbol{\theta}^{*(1)}), U(\boldsymbol{\theta}^{*(2)})$.
        \STATE Apply the optimal circuits and the ground-state energy is estimated by $E_0=c_1\tr[H\rho_1] + c_2\tr[H\rho_2]$
        \STATE Output the ground-state energy $E_0$.
    \end{algorithmic}
\caption{Virtual variational quantum eigensolver (VVQE)}
\label{algo:VVQE}
\end{algorithm}
%%%%%%%%%%%%%%%%%%%%%%%%%%%%%%%%%%%%%%%%%%%%%%%%%%%%%%%%%%%%%%%%%%%%%%%%%%%
\section{Simulating long-range entanglement operations with local operations}\label{sec:simulate non-local gate}

It was proposed in Ref.~\cite{mitarai2021constructing,mitarai2021overhead} that the non-local channel could be simulated by a quasiprobability-based method. Specifically, for a non-local operation $\boldsymbol{\Phi}$, decompose it into the linear combination of local operations and classical communication (LOCC)
\begin{equation}
    \boldsymbol{\Phi} = \sum_i c_i \boldsymbol L_i
\end{equation}
where $c_i$ are the real coefficients, and $\sum_ic_i=1$, and $L_i\in LOCC$. With the decomposition, $\boldsymbol{\Phi}$ can be simulated in a Monte-Carlo manner by sampling $\boldsymbol{L_i}$ with probability proportional $|c_i|$. More concretely, let us define a probabilistic map $\hat{\boldsymbol{\Phi}}$ such that it becomes $\frac{c_i}{|c_i|}\boldsymbol{L_i}$ with probability $p_i=\frac{c_i}{W(\boldsymbol{\Phi})}$ where $W(\boldsymbol{\Phi})=\sum_i|c_i|$, which is known as simulation overhead. Then,
\begin{equation}
    \mathbb{E}[W(\boldsymbol{\Phi}) \hat{\boldsymbol{\Phi}}] = W(\boldsymbol{\Phi}) \cdot \sum_i \frac{|c_i|}{W(\boldsymbol{\Phi})}\frac{c_i}{|c_i|}\boldsymbol{L}_i = \boldsymbol{\Phi},
\end{equation}
which shows that $W(\boldsymbol{\Phi}) \hat{\boldsymbol{\Phi}}$ becomes equal to $\boldsymbol\Phi$ when executed for many times. The non-local gate simulation procedure is the same as the simulation of an unphysical map.

For a two-qubit gate, its non-local part can always be written as Ref.~\cite{kraus2001optimal}
\begin{equation}
    U=\exp\Big[ i\Big(\sum_{\alpha=1}^3\theta_\alpha\sigma_\alpha\ox\sigma_\alpha\Big)\Big] = \sum_{\alpha=0}^3 u_\alpha \sigma_\alpha \ox \sigma_\alpha,
\end{equation}
where $\sigma_0$ is the identity with dimension equal to 2, and $\sigma_1, \sigma_2$ and $\sigma_3$ are Pauli $x,y$ and $z$ operators, respectively. $\theta_\alpha$ is a real parameter, and $u_\alpha$ is a coefficient that is determined from $\{\theta_\alpha\}$. These lead to the following decomposition of channel
\begin{equation}
    \boldsymbol{U}= \sum_\alpha |u_\alpha|^2\boldsymbol{\sigma}_\alpha^{\ox 2} + \sum_{\alpha<\alpha'}(u_\alpha u_{\alpha'}^* + u_{\alpha'} u_\alpha^*)(\boldsymbol{A}_{\alpha\alpha'}^{\ox 2} - \boldsymbol{B}_{\alpha\alpha'}^{\ox 2}) + \sum_{\alpha<\alpha'}i(u_\alpha u_{\alpha'}^* - u_{\alpha'} u_\alpha^*)(\boldsymbol{A}_{\alpha\alpha'}\boldsymbol{B}_{\alpha\alpha'} + \boldsymbol{B}_{\alpha\alpha'}\boldsymbol{A}_{\alpha\alpha'})
\end{equation}
where $\boldsymbol{\sigma}_\alpha$ refers to the Pauli channels, and $\boldsymbol{A}_{\alpha\alpha'} $ and $\boldsymbol{B}_{\alpha\alpha'} $ are the channels that
\begin{equation}
    \boldsymbol{A}_{\alpha\alpha'} (\rho) = \frac{1}{2}(\sigma_\alpha\rho\sigma_{\alpha'} + \sigma_{\alpha'}\rho\sigma_{\alpha}), \quad \boldsymbol{B}_{\alpha\alpha'} (\rho) = \frac{1}{2i}(\sigma_\alpha\rho\sigma_{\alpha'} - \sigma_{\alpha'}\rho\sigma_{\alpha})
\end{equation}
The overhead to simulate the channel $\boldsymbol{U}$ is given in the following
$W(\boldsymbol{U})$
\begin{equation}
    W(\boldsymbol{U}) = 1+\sum_{\alpha\neq \alpha'} (|u_\alpha u_{\alpha'}^* + u_{\alpha'} u_\alpha^*| + |u_\alpha u_{\alpha'}^* - u_{\alpha'} u_\alpha^*|).
\end{equation}
\end{document}